\documentclass[hyperref]{article}
\usepackage{makeidx}
\usepackage[colorlinks=true,citecolor=black]{hyperref}
\usepackage{hyperref}
\usepackage{multirow}
\usepackage{multicol}
\usepackage[dvipsnames,svgnames,table]{xcolor}
\usepackage{graphicx}
\usepackage{epstopdf}
\usepackage{ulem}
\usepackage{hyperref}
\usepackage{amsmath}
\usepackage{amssymb}
\usepackage{enumerate}
\usepackage[a4paper]{geometry}
\usepackage{subfigure}
\usepackage{caption}
\usepackage{tabulary}
\usepackage{makecell}
\makeatother 
\hypersetup{
colorlinks=true,
linkcolor=black
}
\begin{document}
\captionsetup[figure]{labelfont={bf},name={Fig.},labelsep=period}

\title{\bf On the model-checking-based IDS }

\date{}
\author{\large  Weijun ZHU\\
{\sffamily\small School of Information Engineering, Zhengzhou University, Zhengzhou, 450001 China}
}
\maketitle
{\noindent\small{\bf Abstract:}
 How to identify the comprehensive comparable performance of various Intrusion Detection (ID) algorithms which are based on the Model Checking (MC) techniques? To address this open issue, we conduct some tests for the model-checking-based intrusion detection systems (IDS) algorithms. At first, Linear Temporal Logic (LTL), Interval Temporal Logic (ITL) and Real-time Attack Signature Logic (RASL) are employed respectively to establish formula models for twenty-four types of attacks. And then, a standard intrusion set, called Intrusion Set for Intrusion Detection based on Model Checking (ISIDMC) is constructed. On the basis of it, detection abilities and efficiency of the intrusion detection algorithms based on model checking the three logics mentioned above are compared exhaustively.}

\vspace{1ex}

\section{Introduction}
Intrusion detection (ID) is an important network security technique. Its approaches can be mainly classified into two categories: anomaly detection and misuse detection. Compared with the anomaly detection ID methods, the misuse ID ones cannot report unknown types of attacks. However, the latter methods have a comparatively low false positives rate in terms of the known types of attacks. This is due to the principle of misuse detection: Intrusion detection systems (IDS) developers attempt to encode knowledge about attacks. To do this, all the known types of attacks are predefined by using appropriate languages to describe these types, and libraries of attack patterns (called misuse signatures) are established accordingly. The system will monitor the audit log. Once a data stream in the log is found to match with certain attack type, it means that an attack is detected. In other word, misuse detection technique works based on Pattern Matching (PM).

However, this kind of technique suffers from their inherent problems, such as the lack of power of detecting various changing attack patterns \cite{1}\cite{2}. To address this issue, a series of intrusion detection methods based on model checking have been proposed in \cite{1}, \cite{2}, \cite{3} and \cite{4}. Their basic principle can be formulated as follows: (1) using a temporal logic formula to describe an attack pattern, as well as an automaton to record what happened in the audit log, and (2) using a model checking algorithm to check whether the automaton satisfies the formula (i.e. whether the records in log match the attack pattern). The logical operators in temporal formulas can flexibly describe various logical relationships among attack actions.

Compared with the PM-based approaches, the MC-based ones can effectively portray the ever-changing attack patterns \cite{1}\cite{3}. Additionally, the MC-based approaches are better qualified to detect complex intrusion behaviors than the PM-based ones. Pattern matching is usually applicable to detect data inconsistencies. Automata, temporal logic formulas and model checking techniques, in contrast, are applicable to detect behaviors inconsistencies. In other words, the MC-based methods can do more jobs than the PM-based ones since intrusion attacks usually involve not only the comparatively simple data but also complex behaviors \cite{4}. Subsequently, a large and intensive researches concerning model checking were conducted, for aiming to detect attacks with/without malware, such as \cite{19}, \cite{20}, \cite{21} and \cite{22}.

Up to now, we know only a little about comparable detection abilities among the model-checking-based intrusion detection techniques. For example, the MC-based IDS algorithm is the first one for detecting successfully the chop-chop attacks \cite{7}. However, how to evaluate and compare the performances of these MC-based algorithms? It is still an open issue.

To address this issue, we need a benchmark set to evaluate these algorithms. But unfortunately, the existing benchmark set is not available for the model-checking-based IDS algorithms. KDDCUP \cite{5} is a famous benchmark set, but it is only suitable to evaluate data-oriented IDS methods. The model-checking-based methods can deal with behaviors rather than data.

Therefore, motivated by exploring the comparable performance of all of the existing MC-based IDS algorithms, we, in this paper, propose a behavior version of KDDCUP for standard intrusion set and also carry out simulation experiments to evaluate the performance of the three MC-based algorithms, including LTL-based algorithm \cite{1}\cite{6}\cite{7}\cite{8}, ITL-based one \cite{2} and RASL-based one \cite{4}. The experimental results verify that: (1) the RASL-based method has the highest the detection performance, and ITL-based one comes in the second place, followed by LTL-based one; (2) there exits an inverse relationship between the detection performance and the time consumed by these three methods. The law of ``No Free Lunch" works again. That is to say, a uniform standard for evaluation of the model-checking-based IDS algorithms is established. To the best of our knowledge, it is the first one trying to construct a benchmark set for the model-checking-based IDS techniques. As far as we know, it is the first work trying to exhaustively comparing the power and efficiency between all of the existing MC-based IDS algorithms. These are the main contributions of this paper.

The remainder of this paper is organized as follows. Section 2 illustrates three temporal logics. Section 3 establishes some models for twenty-four types of attacks with some temporal logic formulas. Section 4 constructs a log which is a behavior version of KDDCUP. Section 5 presents several groups of experiments and compares the three algorithms based on three temporal logics respectively with regard to the detection capabilities and detection time for intrusion attacks presented in the log. Section 6 draws the conclusions of this paper.

\section{Three temporal logics}
\subsection{LTL}
{\bf Definition 1}, (Syntax of propositional LTL \cite{9}, LTL for short) Let AP be a set of atomic propositions, then:
\begin{itemize}
\item For all $p \in AP$, $p$ is a LTL fomula.
\item If $\varphi$ is a LTL formula, then $\neg \varphi $ is a LTL formula.
\item If $\varphi$ and $\psi$ are LTL formulas, then $ \varphi  \vee \psi $ is a LTL formula.
\item If $\varphi$ is a LTL formula, then $\bigcirc \varphi $ is a LTL formula.
\item If $\varphi$ and $\psi$ are LTL formulas, then $\varphi U\psi $ is a LTL formula.
\end{itemize}
{\bf Definition 2}, (Derived LTL formulas). 

$\diamondsuit \varphi  = trueU\varphi , \Box \varphi  = \neg \diamondsuit \neg \varphi , \varphi  \wedge \psi  = \neg (\neg \varphi  \vee \neg \psi )$

For more details on the formal semantics of the LTL, please refer to \cite{9}.

\subsection{ITL}
{\bf Definition 3}, (Syntax of propositional ITL \cite{10}\cite{11}\cite{12}\cite{13}, ITL for short)

\begin{itemize}
\item For all $p \in AP$, $p$ is a ITL fomula.
\item If $\varphi$, $\psi$ are ITL formulas, the following formulas are ITL ones: $\neg \varphi  ,\varphi  \vee \psi , \bigcirc \varphi ,\varphi ;\psi , \varphi *$
\end{itemize}
{\bf Definition 4}, (Derived ITL formulas).

$\diamondsuit \varphi  = true;\varphi ,\Box \varphi  = \neg \diamondsuit \neg \varphi , \varphi  \wedge \psi  = \neg (\neg \varphi  \vee \neg \psi ), {\varphi _1}||{\varphi _2} = {\varphi _1} \wedge ({\varphi _2};true) \vee {\varphi _2} \wedge ({\varphi _1};true)$
 
The formal semantics of the ITL can be seen in \cite{10}.

\subsection{RASL}
{\bf Definition 5}, RSAL \cite{4}\cite{14}\cite{15} formulas have the following syntax given in Backus-Naur Form: 
\begin{enumerate}[1)]
\item Terms $t:: = T|{T_f}$
\item	 Constraint formulas $\delta :: = {T_f} \le c|{T_f} < c|{T_f} > c|{T_f} \ge c|{\delta _1} \wedge {\delta _2}$
\item Interval formulas \\
$\begin{array}{l}
\varphi :: = p|skip|({\varphi _1},...,{\varphi _m}){\rm{ }}prj{\rm{ }}{\varphi _0}|({\varphi _1},...,{({\varphi _i},...,{\varphi _j})^\Theta } ,...,{\varphi _m}){\rm{ }}prj{\rm{ }}{\varphi _0}|{\varphi _1};{\varphi _2}|\varphi *|{\varphi _1} \vee {\varphi _2}|{\varphi _1} \wedge {\varphi _2}|{\varphi _1}||{\varphi _2}
\end{array}$
\item Timed formulas $\psi :: = \varphi |\delta |{\psi _1} \wedge {\psi _2}|{\psi _1} \vee {\psi _2}|{\psi _1};{\psi _2}$
\end{enumerate}
{\bf Definition 6}, The derived RASL formulas are defined as follows:
\begin{enumerate}[1)]
\item $\bigcirc p:: = skip;p$
\item $more:: = \bigcirc true$ 
\item $empty:: = \bigcirc false$
\item $p{;_I}q:: = (p \wedge {T_f} \in I);q$
\end{enumerate}

The formal semantics of the RASL can be seen in \cite{4}.

\subsection{Some intuitive meaning and a comparison of three logics}
\begin{figure}
		\centering
		\subfigure[a LTL formula]{
			\begin{minipage}[b]{0.8\textwidth}
				\includegraphics[width=\textwidth]{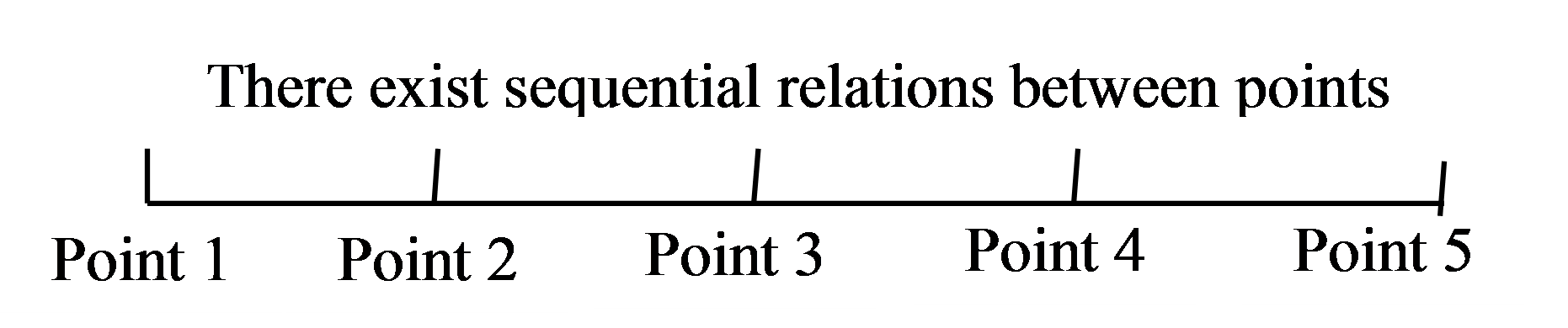} \\
				\label{fig1-1}
			\end{minipage}
		}
		
		\subfigure[ an ITL formula ]
		{
			\begin{minipage}[b]{0.8\textwidth}
				\includegraphics[width=\textwidth]{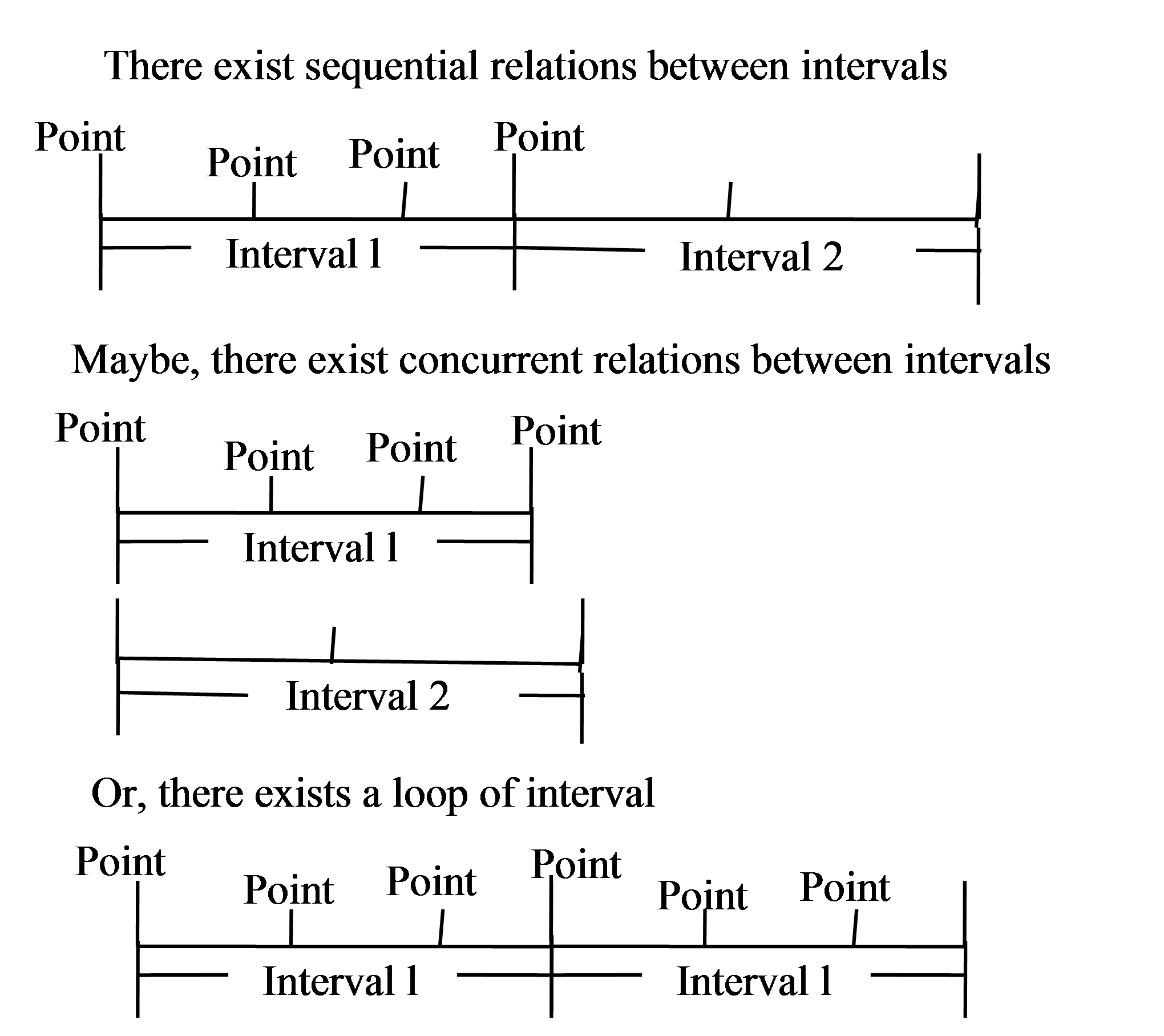} \\
				\label{fig1-2}
			\end{minipage}
		}
		
		\subfigure[a RASL formula]{
			\begin{minipage}[c]{0.8\textwidth}
				\includegraphics[width=\textwidth]{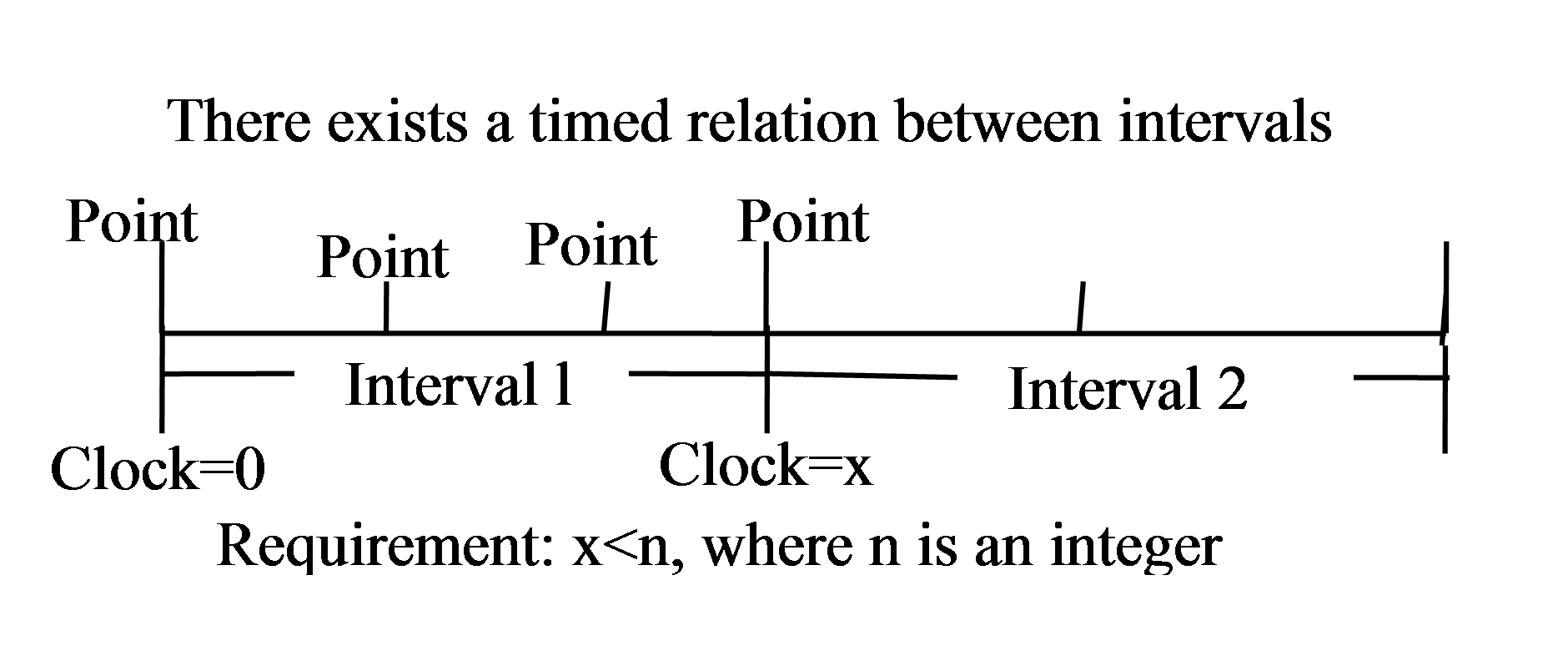} \\
				\label{fig1-3}
			\end{minipage}
		}
		
		\caption{some illustrations for LTL, ITL and RASL}
		\label{fig1}
	\end{figure}

The Fig. \ref{fig1} explains the intuitive meaning of LTL, ITL and RASL. And the Table \ref{tab1} gives the intuitive meaning of the main operators in the three temporal logics.

\section{Constructing attack models with temporal logic formulas}
\subsection{DOS (Denial of Service) attacks }
\subsubsection{Smurf attacks}
\paragraph{Principles}~{}

An attacker on a remote machine sends an Internet Control Message Protocol (ICMP) echo request service, whose destination address is the broadcast address of a network rather than a single host IP address, and its request packet's source IP is the victim's computer IP spoofed by the attacker rather than the attacker’s IP address. As a result, a large number of hosts receive the ICMP echo request packet service, and respond to it by sending a reply to the source IP address. If the number of ICMP packets is large enough, the victim's computer will be flooded with traffic. It results in slowing down the victim's computer to the point where it becomes impossible to work on.

The critical steps can be formulated as follows:
\begin{enumerate}[(1)]
\item send a packet with specific request (such as ICMP echo request) to subnet's broadcast address;
\item and spoof the source address into the being attacked host address.
\end{enumerate}

\paragraph{The formula for Smurf attacks}~{}

According to the principles and its critical steps, we can obtain the atomic propositions for atomic actions in Smurf attacks, as shown in Table \ref{tab2}.

Based on the principles, critical steps and the temporal relationships between the atomic propositions, the following ITL formula for Smurf attacks can be given as follows:
\begin{equation}
\begin{array}{l}
{\varphi _{smurf}} = \neg attacked.send \wedge
{\rm{             }}(true;attacked.recieve)^*
\end{array}
\end{equation}

\subsubsection{Neptune attacks}
\paragraph{Principles}~{}

An attacker sends SYN packets to the destination host with multiple random source address, but does not reply after receiving SYNACK packets from destination host, so that the destination host built up a lot of connection queues for these source hosts. Since the destination host has not received ACK packets, it has to consume a growing systemic resource to maintain these queues, resulting in the disability to provide service for normal requests.

The critical steps can be formulated as follows: 
\begin{enumerate}[(1)]
\item an attacker sends a SYN packet to the attacked host; 
\item after the attacked host returns a SYNACK message, the attacker does not return any ACK packet.
\end{enumerate}
\paragraph{The formula for Neptune attacks}~{}

According to the critical steps and the principles, we obtain the atomic propositions for atomic actions in Neptune attacks, as shown in Table \ref{tab3}.

In line with the critical steps, the principles and the temporal relationships between the atomic propositions, LTL formula for Neptune attacks can be defined as follows:
\begin{equation}
\begin{array}{l}
{\varphi _{nept}} = (attacked.recieve.SYN \wedge 
{\rm{       }}\diamondsuit attacked.send.SYNACK \wedge 
{\rm{      }}\Box(\neg attacked.recieve.ACK))
\end{array}
\end{equation}

\subsubsection{LAND (Local Area Network Denial) attacks}
\paragraph{Principles}~{}

An attacker sends a spoofed TCP SYN packet with the attacked host's IP address to an open port as both source and destination. This causes the attacked host to reply to itself endlessly., which significantly reduces system performance.

The critical steps can be formulated as follows:
\begin{enumerate}[(1)]
\item an attacker sets both source address and destination address of a packet as the attacked host address; 
\item  he/she sends this packet to the attacked host via IP spoofing.
\end{enumerate}
\paragraph{The formula for Land attacks}~{}

As stated in the critical steps and the principles, we obtain the two atomic propositions, as shown in Table \ref{tab4}.

According to the principles and the logical relationships between the atomic propositions, we obtain the following propositional formula for LAND attacks:
\begin{equation}
{\varphi _{land}} = attacked.recieve \wedge p
\end{equation}

\subsubsection{Teardrop attacks}
\paragraph{Principles}~{}

When an IP data packet is transferring on the network, it will be divided into smaller fragments. In the IP header, one of the fields is the fragment offset field, which indicates the starting position of the data contained in a fragmented packet. An attacker can launch Teardrop attack through sending two (or more) fragments. The offset of the first fragment is 0, and the length is N. The offset of the second one is smaller than N. As a result, the two packets overlap, and the server has to attempt to reassemble the packet, and TCP / IP stack needs to allocate unusually vast resources, resulting in a lack of system resources or even restart the machine.

The critical steps can be formulated as follows: 
\begin{enumerate}[(1)]
\item the offset of the first data segments is 0, and its length is N; 
\item the offset of the second one is smaller than N.
\end{enumerate}
\paragraph{The formula for Teardrop attacks}~{}

According to the critical steps and the principles, we obtain the atomic propositions for atomic actions in Teardrop attacks, as shown in Table \ref{tab5}.

According to critical steps, the principles and the temporal relationships between the atomic propositions, we obtain the following LTL formula for Teardrop attacks:
\begin{equation}
\begin{array}{l}
{\varphi _{tear}} = m1.FragmentOffset = 
{\rm{           }}0 \wedge m1.TotalLength = 
{\rm{          }}N \wedge \diamondsuit (m2.FragmentOffset < N)
\end{array}
\end{equation}

\subsubsection{Pod attacks}
\paragraph{Principles}~{}

According to TCP / IP specification, the maximum length of a packet is 65536 bytes. However, the superposition of multiple fragments of one packet can exceed 65536 bytes. The host is being received a Ping of Death attack, which will cause the host downtime, when it receives a packet whose length is greater than 65536 bytes.

The key issue is to detect whether the length of received packet is greater than 65536 bytes or not.

\paragraph{The formula for Pod attacks}~{}

According to the principles, we obtain the only one atomic proposition for atomic action in Pod attacks, as shown in Table \ref{tab6}.

According to the principles, the propositional formula for Pod attacks can be given by:
\begin{equation}
{\varphi _{pod}} = m.size > 65536
\end{equation}

\subsubsection{Mailbomb Attacks}
\paragraph{Principles}~{}

An attacker sends huge volumes of duplicate mails to the same mail address in a very short time by using some specially prepared mail sending software. These mails may come with a large attachment. The mailbox will become unusable, when it confronts with a large number of mail packets and its storage capacity is limited. Under this scenario, if the mailbox is not cleaned up in time, resulting in mailbox overflow, the subsequent normal mails will be returned. 

If the mail server cannot work properly for a long time, it will cause that the normal user's E-mail cannot be received and delivered.

The situation may deteriorate since the large number of returned mails will lead to more system and network resource consumption, the obstruction of traffic and other security issues on the mail server. 

Mail bomb utilizes the Simple Mail Transfer Protocol (SMTP), the mail receiving protocol (POP3), and the Internet Mail Extension Protocol (MIME). Attackers often use some special tools to quickly send a lot of spam. One can fake the mail delivery address, but the source IP address cannot be faked. As a result, the attackers use the same IP address send a lot of concurrent SMTP links.

When a source IP address sends large mails to the same destination IP address within the specified time interval (the default is 0.01 seconds), one can determine that the host is under a Mail bomb attack.

Thus, the atomic actions and the properties of a mail bomb attack can be portrayed as follows.
\begin{enumerate}[(1)]
\item Starting with each mail record, and search ten records chronologically.
\item Determining whether all of the time intervals of the ten records are no more than 0.01 seconds or not. If the answer is yes, it indicates that the mail host is under a Mail bomb attack.
\end{enumerate}

\paragraph{The formula for Mailbomb Attacks}~{}

According to the principles of Mail bomb attacks, we obtain the atomic propositions for the atomic actions and the properties of it, as shown in Table \ref{tab15}. Therefore, the formula for a mail bomb attack can be given as follows.

\begin{equation}
\begin{array}{l}
({p_1}{;_{x < 0.01}}{p_2}{;_{x < 0.01}}{p_3}{;_{x < 0.01}}{p_4}{;_{x < 0.01}}{p_5}{;_{x < 0.01}}
{p_6}{;_{x < 0.01}}{p_7}{;_{x < 0.01}}{p_8}{;_{x < 0.01}}{p_9}{;_{x < 0.01}}{p_{10}})^*
\end{array}
\end{equation}

\subsubsection{UDPstorm Attacks}
\paragraph{Principles}~{}

An attacker prepares enough puppet hosts which send a large number of UDP packets to random ports on a remote host. For the large number of UDP packets, the victimized host will be forced into sending many ICMP packets. These packets will further exhaust the system bandwidth, eventually leading the target host to be unreachable by other clients. And such attack not only damages the target host, but also causes the congestion of intermediate routers or switching equipment.

There is a buffer receiving packets inside the UDP port. When it is receiving some data packets, UDP port will verify to the destination port number which is included in the received packet. If the port number matches the one of local host, the packet will be sent into the receiving queue. Otherwise, the packet will be discarded, and an ICMP packet will be sent to the source, indicating the destination has been denied. When the local host receives a UDP packet and it finds that the received port number does not match the currently used port, it will send an ICMP packet indicating the unreachable destination to the faked source address. If an attacker sends enough UDP packets, it will render the associated network device to be brought down.

In short, the target host receives a large number of UDP packets sent from all the hosts in the subnet, but the destination ports of these packets do not match the currently used ports. And the target host sends ICMP error messages to the source address.
Thus, the atomic actions and the properties of an Udpstorm attack can be portrayed as follows.

Thus, the atomic actions and the properties of an Udpstorm attack can be portrayed as follows.
\begin{enumerate}[(1)]
\item The target host receives the UDP packet sent from the host in the subnet.
\item The port number included in the packet does not match the port number of the currently target host.
\end{enumerate}

\paragraph{The formula for UDPstorm Attacks}~{}

According to the principles described above, we obtain the atomic propositions for the atomic actions and the properties in an UDPstorm attack, as shown in Table \ref{tab16}. Therefore, the formula for an Udpstorm attack can be given as follows.

\begin{equation}
\forall \;i[(attacked.receive.i) \wedge (attacked.port \ne i.udp.port)]
\end{equation}
\subsubsection{Apache Attacks}
\paragraph{Principles}~{}

An attacker can launch a remote attack by exploiting the vulnerability of Apache Server (one of the most popular Web servers), causing the server host system's resources to be exhausted and unable to respond to normal requests. Furthermore, an insecure policy occurs in the Apache HTTP Server when this Sever is processing a Range option in an HTTP request packet. An attacker can use a larger Range Header Threshold (default 5) to construct as many requests for different sized page files as possible. Thus, the server's resources, i.e., memory and CPU, are consumed tremendously. The attacker can also encapsulate the request by loading the ``Accept-Encoding" option, which will further enhance the effect of the attack. As a result, Apache cannot respond to normal users' requests, because the resource of server will be exhausted.

Summarily, the target host receives an HTTP request, where Range is greater than 5. And these packets carry some Accept-Encoding information.

Thus, the atomic actions and the properties of an Apache attack can be portrayed as follows.
\begin{enumerate}[(1)]
\item The target host receives an HTTP request.
\item The value of Range in the request packet is greater than 5.
\item The request packet carries Accept-Encoding information.
\end{enumerate}

\paragraph{The formula for Apache Attacks}~{}

According to the principles as above, we obtain the atomic propositions for the atomic actions and the properties in an Apache attack, as shown in Table \ref{tab17}. Therefore, the formula for an Apache attack can be given as follows.
\begin{equation}
\begin{array}{l}
p \wedge (attacked.receive.http.range > 5B) \vee 
\left( {attacked.receive.http.accept - encoding = 1} \right)
\end{array}
\end{equation}

\subsection{Probing attacks}
\subsubsection{IP sweep}
\paragraph{Principles}~{}

When a source IP address sends 10 ICMP packets to different hosts within the specified time interval (default is 0.01 seconds), by which the network firewall can detect that an address sweep is conducted.

The key steps can be formulated as follows. 
\begin{enumerate}[(1)]
\item starting with every record (i.e., an ICMP packet), and sequentially check ten records.
\item For every ten records, checking whether the time interval between the first record and the last one is 0.01 second or not. If all of the time intervals are more than 0.01 second, then no IP sweep occurs. 
\item checking the addresses of destination hosts of these ten records. If they are all different from each other, we can confirm that an IP sweep attack has been conducted.
\end{enumerate}

\paragraph{The formula for IP sweep}~{}

According to the critical steps and the principles as above, we obtain the atomic propositions for atomic actions in IP sweep, as shown in Table \ref{tab7}.

According to critical steps, the principles and the temporal relationships between the atomic propositions, we obtain the following RASL formula for IP sweep:

\begin{equation}
\begin{array}{l}
{\varphi _{ipsw}} =  \neg \Box ((p1;p2;p3;p4;
{\rm{           }}p5;p6;p7;p8;p9){;_{x \ge n}}p10)
\end{array}
\end{equation}

\subsubsection{Port scan}
\paragraph{Principles}~{}

When a source IP address sends IP packets with TCP/SYN fragment to 10 different ports which have the same destination IP address within the specified time interval (default is 0.01 seconds), the network firewall can identify that a port scan attack occur on the host system.

The critical steps can be formulated as follows.
\begin{enumerate}[(1)]
\item starting with every record (i.e., TCP/SYN packet), and sequentially check ten records.
\item for every ten records, checking whether the time interval of the first record and the last one is no less than 0.01 seconds or not. If yes, no Port scan occurs.
\item checking the addresses of destination hosts of these ten records. If they are all different from each other, we can further confirm that a Port scan has been conducted.
\end{enumerate}

\paragraph{The formula for Port scan}~{}

According to the critical steps and the principles as mentioned above, we obtain the atomic propositions for atomic actions in Port scan, as shown in Table \ref{tab8}.

On the basis of the key steps, the principles and the temporal relationships between the atomic propositions as above, we obtain the following RASL formula for Port scan:

\begin{equation}
\begin{array}{l}
{\varphi _{port}} = \neg \Box ((q1;q2;q3;q4;
{\rm{          }}q5;q6;q7;q8;q9){;_{x \ge n}}q10)
\end{array}
\end{equation}

\subsubsection{Nmap}
\paragraph{Principles}~{}

Nmap (Network mapper) mainly consists of four parts: host discovery, port scanning, service detection, and operating system scan.

The critical steps can be formulated as follows.
\begin{enumerate}[(1)]
\item host discovery: Send four kinds of detection packets: ICMP echo request, a TCP SYN packet to port 443, a TCP ACK packet to port 80, and an ICMP timestamp request.
\item port scanning: it contains the following actions.

TCP SYN scanning: An attacker sends a SYN packet to the destination port. If a SYN/ACK is received, it indicates that the port is listening (i.e., open). If a RST packet is returned, it indicates that the port is closed. If no response is returned, it shows the port is shielded.

TCP ACK scanning: An attacker sends an ACK packet to destination host. If a RST packet is returned, it indicates that the port has been not blocked by a firewall. Otherwise, this port has been shielded.

TCP FIN scanning: An attacker sends a TCP/FIN packet to destination host. If a RST packet is returned, it indicates that the port has been closed. Otherwise, this port has been open or shielded.

TCP Xmas scanning: An attacker sends a Xmas tree packet to destination host. If a RST packet is replied, this goes to show that the port has been closed. Otherwise, this port has been opened or shielded. The Xmas tree packet is the TCP packet whose FIN URG PUSH of the flags field is set to 1.

TCP NULL scanning: An attacker sends a NULL packet to destination host. If a RST packet is returned, it indicates that the port has been closed. Otherwise, this port has been open or shielded. The NULL packet is the TCP packet, all of whose flags fields are set to 0.

UDP scanning: An attacker sends a UDP packet to destination host. If message ``ICMP port unreachable" is returned, it shows that the port has been closed. If no reply is returned, the UDP port has been open or shielded.

\item version detection: It is used to confirm the applications and version information on the open ports of the target host.
\item operating system detection: first, an open port and a closed port are selected, respectively, and some designed TCP/UDP/ICMP packets are sent to these ports. And then, a system fingerprint is generated according to the reply packets. Besides that, the generated fingerprint is contrasted with the nmap-os-db one to find the matching system. If the match fails, the possible systems are listed.
\end{enumerate}

\paragraph{The formula for nmap}~{}

According to the critical steps and the principles mentioned above, we obtain the atomic propositions for atomic actions in nmap, as shown in Table \ref{tab9}.

Based on critical steps, the principles and the temporal relationships between the atomic propositions, we obtain the following ITL formula for nmap:
\begin{equation}
\begin{array}{l}
{\varphi _{nmap}} = ({\varphi _1};{\varphi _2};{\varphi _3};{\varphi _4});anyotherattack\_formula\\
where,\\
{\varphi _1} = attacked.recieve.ICMP - echo - request \vee (attacked.recieve.TCPSYN \wedge port = 443) \vee \\
{\rm{ }}(attacked.recieve.TCPACK \wedge port = 80) \vee attacked.recieve.ICMP - timestamp - request\\
{f_1} = attacked.recieve.SYN,\\
{f_2} = attacked.recieve.ACK,\\
{f_3} = attacked.recieve.TCPFIN,\\
{f_4} = attacked.recieve.TCP.flags.FINURGPUSH = 1,\\
{f_5} = attacked.recieve.TCP.flags = 0,\\
{f_6} = attacked.port.UDP.recieve.ICMP,\\
{\varphi _2} = {f_1} \vee {f_2} \vee {f_3} \vee {f_4} \vee {f_5} \vee {f_6},\\
{f_7} = Exclusionlist.check.status.open,\\
{f_8} = Exclusionlist.check.status.openfiltered,\\
{f_9} = port.TCP.TCPconnect,\\
{f_{10}} = port.UDP.recieve.nmapserviecesprobes,\\
{\varphi _3} = ({f_7} \vee {f_8}) \wedge \diamondsuit ({f_9} \vee {f_{10}}),\\
{f_{11}} = port.open.recieve.TCP,\\
{f_{12}} = port.open.recieve.UDP,\\
{f_{13}} = port.open.recieve.ICMP,\\
{f_{14}} = port.closed.recieve.TCP,\\
{f_{15}} = port.closed.recieve.UDP,\\
{f_{16}} = port.closed.recieve.ICMP,\\
{\varphi _4} = ({f_{11}} \vee {f_{12}} \vee {f_{13}}) \wedge ({f_{14}} \vee {f_{15}} \vee {f_{16}}),
\end{array}
\end{equation}

\subsubsection{Satan}

\paragraph{Principles}~{}

There are three levels of scans depending on the network circumstance, which are low-grade scan, normal scan and grievous scan.
The main steps can be formulated as follows: (1) or (2) or (3), where:

\begin{enumerate}[(1)]
\item A low-grade scan starts from Domain Name System (DNS), rpc or portmap. A DNS scan uses the nslookup (A kind of network administration command-line tool on Unix-like which can query to obtain domain name or IP address mapping) to gather more information about the target host. And, rpc/portmap scans work by using showmount programs or portmap programs.
\item Normal scans contain all content of low-grade scanning, besides that, it also scan fingered, TCP and UDP services.
\item Grievous scans not only contain the entire contents of Low-grade scans and Normal scans, but also scan very active services.
\end{enumerate}

\paragraph{The formula for Satan}~{}

According to the critical steps and the principles as stated above, we obtain the sub-formulas in Satan, as shown in Table \ref{tab10}.

According to critical steps, the principles and the temporal relationships between the sub-formulas, we obtain the following ITL formula for Satan:

\begin{equation}
\begin{array}{l}
{\varphi _{sata}} = {\varphi _1} \vee {\varphi _2} \vee {\varphi _3}\\
where,\\
ITL{f_1} = attacked.nslookuped\_program,\\
ITL{f_2} = attacked.portmapped\_program,\\
ITL{f_3} = attacked.showmount\_program,\\
{\varphi _1} = ITL{f_1} \vee ITL{f_2} \vee ITL{f_3},\\
ITL{f_4} = attacked.fingered.scanned\_program,\\
ITL{f_5} = attacked.TCP.scanned\_program,\\
ITL{f_6} = attacked.UDP.scanned\_program,\\
{\varphi _2} = (ITL{f_1} \vee ITL{f_2} \vee (ITL{f_2} \wedge ITL{f_3})) \wedge {\rm{ }}(ITL{f_4} \vee ITL{f_5} \vee ITL{f_6}),\\
ITL{f_7} = attacked.activeservices.scanned\_program,\\
{\varphi _3} = {\varphi _2} \wedge ITL{f_7}\\

\end{array}
\end{equation}
\subsubsection{Mscan}

\paragraph{Principles}~{}

Mscan searches exhaustively all of the IP address in the whole domain in order to find the running hosts and their vulnerabilities. This program does not need root privileges. And when Mscan runs on Linux, it can effectively scan some vulnerabilities on the remote target hosts.

The main points are listed as follows.
\begin{enumerate}[(1)]
\item port scanning

TCP SYN scanning: for port 113, if a SYN / ACK reply is received, the port's state is open; if a RST packet is received, the port's state is close; if no reply is received, the port's state is shield.

TCP connect scanning: for port 21, ``API connect" of system is used to start a connection to the target host's port. If the connection fails, the port's state is close.

TCP ACK scanning: for port 389, it receives a RST packet after an ACK packet is sent to the target host, indicating that the port is normal. If the RST packet is not received, the port's state is shield.

TCP FIN scanning: for port 443, a FIN packet is sent to the destination port, which will return a RST to all closed ports.

TCP Xmas scanning: for port 443, a FIN packet is sent to the destination port, which will return a RST, i.e., a flag, to all closed ports, where ``FIN URG PUSH" in the RST is set to 1.

TCP NULL scanning: for port 443, a FIN packet is sent to the destination port, which will return a RST, i.e., a flag, to all closed ports, where all flags are set to 0.

UDP scanning: check the target's UDP port. An UDP port packet is sent via the port 971.If it receives`` ICMP port unreachable", the target port's state is close. Otherwise, the target port's state is open.

\item  The port mapper provides a running RPC service

In order to discover a port that a given RPC service can be reached, the client will establish a TCP connection on port 135, requesting a port that is assigned to a particular RPC service.
\item NFSD provides a table of contents

NFSD is a process which can remotely access network file system. And this process can handle some requests on file sharing from client machines. NFSD will work when the output list exists. The attacker uses an exportfs command to detect it. If nothing or something wrong is returned, it shows that this process's state is close or shield.

\item Samba or netbios provides a shared table of contents

The service programs called amba and netbios provide some accesses to shared resources for the client machines on Windows or UNIX OS.
\item use finger to verify whether or not the default account exists

A server system stores the details of each user machines. And the system will return these detailed data to the query, when a finger query occurs.

\item detection of operating systems

TCP / UDP / ICMP packets are sent to the target host via some open ports. The packets returned from the host are compared with the finger database in the system. In this way, the probable system is determined according to the principle of probability matching.

\end{enumerate}

Thus, the atomic actions of a Mscan attack can be portrayed as follows.
\begin{enumerate}[(1)]

\item There are seven ways of port scanning: TCP SYN scanning, TCP connect scanning, TCP ACK scanning, TCP FIN scanning, TCP Xmas scanning, TCP NULL scanning and UDP scanning;
\item The portmapper provides a running RPC service.
\item NFSD provides a table of contents.
\item Samba or netbios provides a shared table of contents.
\item use finger to determine whether or not the default account exists.
\item detection of operating systems.

\end{enumerate}

\paragraph{The formula for Mscan}~{}

According to the principles, we obtain the (atomic) propositions for the (atomic) actions and the properties in a Mscan attack, as shown in Table \ref{tab18}. Therefore, the formula for a Mscan attack can be given as follows.

\begin{equation}
\begin{array}{l}
\Box [(attacked.receieve.SYN \wedge port = {\rm{113}}) \vee (attacked.receieve.SYN \wedge port = {\rm{21}})\\
 \vee (attacked.receieve.SYN \wedge port = {\rm{389}}) \vee (attacked.receieve.SYN \wedge port = {\rm{443}}) \vee \\
(attacked.receieve.TCP.flags.FINURGPUSH = {\rm{1}} \wedge port = {\rm{443}}) \vee \\
(attacked.receieve.TCP.flags = 0 \wedge port = {\rm{443 }}) \vee (attacked.port.UDP.receive.ICMP\\
 \wedge port = {\rm{971}}) \vee (attacked.receieve.TCP.{\rm{ }}portmapper \wedge port = {\rm{135}}) \vee (attacked.receieve.nfsd.exportfs)\\
 \vee (attacked.receieve.samba \vee attacked.receieve.netbios) \vee (attacked.receieve.finger \wedge port = {\rm{79}}) \vee \\
((port.open.receive.TCP \vee port.open.receive.UCP \vee port.open.receive.ICMP) \wedge (port.closed.receive.TCP\\
 \vee port.{\rm{ }}closed.receive.UCP \vee port.closed.ICMP))]
\end{array}
\end{equation}

\subsection{U2R(User-to-Root)attacks}
\subsubsection{Buffer Overflow}

\paragraph{Principles}~{}

A attacker can deliberately arrange code in the address space of program. And then, he/she makes the program jumps to the address space and execute through proper initialization of registers and memory. There are two ways to arrange appropriate code in the address space of program: implantation and use of the existing code. For the former way, the attacked host receives a string containing attack sequence which can run on the attacked hardware platforms. The latter way works by using the current code: modify the parameters of the existing codes and then execute the codes.

The key steps can be formulated as follows. 
\begin{enumerate}[(1)]
\item Arrange appropriate code in the address space of program. 
\item Make the program jumps to the address space and execute program.
\end{enumerate}

\paragraph{The formula for Buffer Overflow}~{}

According to the critical steps and the principles, we obtain the atomic propositions or sub-formulas in Buffer Overflow, as shown in Table \ref{tab11}.

According to critical steps, the principles and the temporal relationships between the atomic propositions/sub-formulas, we obtain the following ITL formula for Buffer Overflow:

\begin{equation}
\begin{array}{l}
{\varphi _{buff}} = attacked.recieve.string \wedge ITL{f_1}^* \vee code.modified;ITL{f_2}^*\\
where,\\
ITL{f_1} = string\_program,\\
ITL{f_2} = code.execute\_program.
\end{array}
\end{equation}

\subsubsection{Rootkit}

\paragraph{Principles}~{}

An attacker replaces ps, ls, netstat and df programs with some programs in Rootkit. Thus, the system administrator cannot find the track through these tools. After that, the attacker clears the syslog with the log cleanup tool in order to eliminate his/her tracks.
The main steps can be formulated as follows. 
\begin{enumerate}[(1)]
\item The monitoring programs in a system are replaced. 
\item Syslog is modified or deleted.
\end{enumerate}

\paragraph{The formula for Rootkit}~{}

On the base of the critical steps and the principles, we obtain the sub-formulas in Rootkit, as shown in Table \ref{tab12}.

According to critical steps, the principles and the temporal relationships between the sub-formulas, we obtain the following ITL formula for Rootkit:

\begin{equation}
\begin{array}{l}
{\varphi _{root}} = ITL{f_1};(ITL{f_2} \vee ITL{f_3})\\
where,ITL{f_1} = code.modified\_program,\\
ITL{f_2} = syslog.modified\_program,\\
ITL{f_3} = syslog.delete\_program.
\end{array}
\end{equation}

\subsubsection{Httptunnel}

\paragraph{Principles}~{}

Some hosts and a server are located in a LAN. And a host and the server are inside and outside a firewall, respectively. It is unlikely that the two machines communicate with each other because the firewall shields all ports except for port 80.One can establish a two-way channel using the software virtualization technique, which loads the protocol that is prohibited by the firewall into a legal protocol. And it uses the way of port conversion to transform a prohibited port into an authorized port. As a result, both sides of the communication break through the restrictions of the firewall. This is the firewall penetration technique.

The HTTP tunnel is established in a circumstance of limited communication in networks. A client can access its server without being restricted by the firewall, by establishing a two-way channel between the server and the client. It is worth noting that the client is inside the firewall, and the configuration of the firewall allows only data packets on the 80th port to go through. In contrast, the server is outside the firewall, and it is open. Thus, it seems impossible for the client to visit the POP3 service on the server. However, Httptunnel attacks can do it. The Httptunnel includes the following two procedures, htc on the client and hts on the server. The connection process is portrayed as follows.

Step 1: the Httptunnel run the htc on the client host, in order to listen on each unused port $(1024<Port number<65535)$. And this port is employed to encapsulate into port 80. As a result, a legitimate request on HTTP connection occurs.

Step 2: the Httptunnel run the hts on the server. And it transforms all received packets on port 80 into the corresponding applied port. The latter port number is in conformance with the port number from the sender.

Step 3: The client sends an HTTP request packet.

The main points are listed as follows.

As shown in the log file, the htc runs on the client host, and all the packets to be sent the external network on the ports between 1024 and 65535 are encapsulated into the packets on port 80. What is more, it sends the encapsulated HTTP request through the firewall to the external networks.

Thus, the atomic actions and the properties of a Httptunnel attack can be portrayed as follows.

\begin{enumerate}[(1)]
\item the htc runs on client host;
\item All the packets on the ports between 1024 and 65535 are encapsulated into the packets on port 80.
\item The htc sends the encapsulated HTTP request through the firewall to the external networks.
\end{enumerate}

\paragraph{The formula for Httptunnel}~{}

According to the principles, we obtain the atomic propositions for the atomic actions and the properties in a Httptunnel attack, as shown in Table \ref{tab19}. Therefore, the formula for a Httptunnel attack can be given as follows.
\begin{equation}
client.htc \wedge \mathop \forall \limits_{i = 1024}^{65535} packets.port.i = 80 \wedge \diamondsuit (client.send.http)
\end{equation}

\subsubsection{Xterm}

\paragraph{Principles}~{}

Xterm is a standard virtual terminal on the X Window System. A user can simultaneously open many Xterms on a single monitor, and each terminal emulator can provide independent input and output. There exist some vulnerabilities in local buffer of Xterm. Thus, a local attacker can use these vulnerabilities to obtain local root privileges. A typical Xterm attack is portrayed as follows.
\begin{enumerate}[(1)]
\item The program does not check the boundary values for the user's codes. An attacker sends some Escape codes to change the size of a window. When the upper bound is greater than 65535, all memory are exhausted, resulting in Xterm crash.
\item An attacker modifies some information on banner in processes of FTP or TELNET.
\item The Attacker modifies the records in a system log. The attack maybe happens when the administrator monitors these log files in xterm / rxvt.
\item Xterm creates a system log. If an attacker creates a symbolic link between checking whether the file can be written and opening this file with $O\_WRONLY | O\_APPEND$ permission, a competitive relationship occurs. Thus, the attacker gains the root privilege.
\item  There exist some vulnerabilities when Xterm filters some sequential symbols in loop semantics. If the attacker send a command with $`` \$  echo-e" \ eP0; 0 | 0A / 17 \ x9c $", Xterm will be trapped in a vicious cycle, and use up the resource of the system and make the system breakdown.
\end{enumerate}

The key points for detection of this kind of attacks are: an attack occurs if one of the above situations happens.

Thus, the atomic actions and the properties of an Xterm attack can be portrayed as follows.
\begin{enumerate}[(1)]
\item The target host receives an Escape command, and the size of window is greater than 65,535;
\item The information on banner of the target host is modified on FTP or TELNET;
\item The systematic log on target host is modified;
\item $O\_WRONLY | O\_APPEND$ of the system log, there is a symbolic link between the two permissions.
\item The target host has received a malicious loop escape character $`` \$ echo-e`` \ eP0; 0 | 0A / 17 \ x9c $" code.
\end{enumerate}

\paragraph{The formula for Xterm}~{}

According to the principles, we obtain the atomic propositions for the atomic actions and the properties in an Xterm attack, as shown in Table \ref{tab20}. As a result, the formula for an Xterm attack can be given as follows.

\begin{equation}
\begin{array}{l}
(Attack.receive.escape.threshold \vee 
Banner.modifed.FTP \\
 \vee Banner.modifed.TELNET 
 \vee Syslog.modifed 
\vee Syslog.{\rm{ }}Symlinked \vee p)^*
\end{array}
\end{equation}
\subsection{R2L (Remote-to-Login)attacks}

\subsubsection{Warezmaster}

\paragraph{Principles}~{}

Warezmaster exploits a system bug in a FTP server. In general, guest users are never allowed to write at a FTP server, therefore they cannot upload files on the server. These guest accounts on most public FTP servers are designed for downloading data. Thus, anyone can access the FTP server by using guest accounts. This type of attack takes place when a FTP server has given write permissions by mistake to users on the system. Thus, any user can login and upload files.

The critical steps can be formulated as follows.
\begin{enumerate}[(1)]
\item an attacker logins to the server with guest account. 
\item the attacker creates a hidden directory and uploads ware-z to the server.
\end{enumerate}

\paragraph{The formula for Warezmaster}~{}

According to the critical steps and the principles, we obtain the sub-formulas in Warezmaster, as shown in Table.\ref{tab13}.

According to critical steps, the principles and the temporal relationships between the sub-formulas, we obtain the following ITL formula for Warezmaster:

\begin{equation}
\begin{array}{l}
{\varphi _{warem}} = ITL{f_1};ITL{f_2};ITL{f_3}\\
where,ITL{f_1} = account.guest.login\_program,\\
ITL{f_2} = hiddendirectory.created\_program,\\
ITL{f_3} = uploadwarez\_program.
\end{array}
\end{equation}

\subsubsection{Warezclient}

\paragraph{Principles}~{}

Warezclient attacks can be launched by any legal user during a FTP connection after a warezmaster attack has been executed. During a process of warezclient attack, a user downloads the illegal ware-z software that has been uploaded through a successful warezmaster attack.

The critical steps can be formulated as follows. 
\begin{enumerate}[(1)]
\item Carrying out warezmaster an attack.
\item Downloading illegal software ware-z.
\end{enumerate}

\paragraph{The formula for Warezclient}~{}

According to the critical steps and the principles, we obtain the sub-formulas in Warezclient, as shown in Table \ref{tab14}.

According to critical steps, the principles and the temporal relationships between the sub-formulas, we obtain the following ITL formula for Warezclient:

\begin{equation}
\begin{array}{l}
{\varphi _{warez}} = {\varphi _{warem}};ITL{f_1}\\
where,ITL{f_1} = downloadwarez\_program.
\end{array}
\end{equation}

\subsubsection{Ftp\_write}

\paragraph{Principles}~{}

In general, many FTP servers allow to be accessed by anonymous users (guest account). And a guest account can only access the files in a specific directory. An attacker can place a fake file called ``.rhosts" in the home directory, which allows an attacker to open a remote session called rlogin. Once the above operation is done successfully, the attacker can easily log in to the FTP server without any password via rlogin.Thus, the unprotected files are able to arbitrarily run, read or written by the attacker.

The main points are listed as follows. An attacker places a fake file called ``.rhosts" in the home directory. And the FTP server opens the process of rlogin.

Thus, the atomic actions of a Ftp\_write attack can be portrayed as follows.

\begin{enumerate}[(1)]
\item Create a new fake file (named ``.rhosts") on the home directory on the FTP server.
\item The FTP server opens the process of rlogin.
\end{enumerate}

\paragraph{The formula for Ftp\_write}~{}

According to the principles, we obtain the atomic propositions for the atomic actions and the properties in a Ftp\_write attack, as shown in Table \ref{tab21}. Therefore, the formula for a Ftp\_write attack can be defined as follows

\begin{equation}
\begin{array}{l}
[(attacked.{\rm{ }}create.file \wedge (file.p = rhosts)) \to 
\diamondsuit (attacked.open.rlogin)]^*
\end{array}
\end{equation}

\subsubsection{Phf}

\paragraph{Principles}~{}

The essence of Phf attacks is not to comply with the rules and some violations of string constraints in a page. An attacker can run arbitrary malicious code on a WEB server in the guise of a privileged user with administrative access controls, by exploiting the given vulnerability.

Normally, Phf scripts are employed to provide some examples for illustrating functions on the WEB. If these scripts with regard to the input format have some vulnerabilities, an attacker can send some random data which are seemingly safe to the scripts, according to a special pre-defined format. Thus, the attacker can tamper with the password of the WEB server, and get an access to command lines or shell of the target system.

The main points are listed as follows.

If the target server receives an HTTP request with a newline ( ``$\verb|\|  {} n$", 0x0a) and the target server executes the Xterm command, or the target server generates a reverse telnet connection on port 90 and port 25, one can determine the system has suffered a Phf attack.

Thus, the atomic actions and the properties of a Phf attack can be portrayed as follows.

\begin{enumerate}[(1)]
\item The target server receives an HTTP request.
\item This HTTP request contains the conversion character, i.e., ``$\verb|\|  {} n$" and hexadecimal 0x0a;
\item The target server executes an Xterm command.
\end{enumerate}

\paragraph{The formula for Phf}~{}

According to the principles, we obtain the atomic propositions for the atomic actions and the properties in a Phf attack, as shown in Table \ref{tab22}. As a result, the formula for a Phf attack can be given as follows
\begin{equation}
\begin{array}{l}
Attacked.receive.http \wedge p \wedge \bigcirc (Attack.xterm;
Attacked.te\ln et \wedge port = 25;
Attacked.te\ln et \wedge port = 90)
\end{array}
\end{equation}

\subsubsection{Imap}

\paragraph{Principles}~{}

A client can use Internet Message Access Protocol (IMAP) to obtain information on mails and download mails from the mail server. This is the major function of IMAP. 

An IMAP server typically listens on port number 143 in the TCP / IP protocol. One can operate mails on the server via the mail client, without downloading these mails.

In the most EMAIL clients supporting the IMAP protocol, its ``literal" value is not been checked carefully with regard to its boundary buffer. A remote attacker can exploit this vulnerability to make a client-side’s buffer overflow. Thus, he or she can get a permission to execute arbitrary commands on the system.

The main points are listed as follows.

An attacker utilizes the false boundary condition to let the value of ``literal" set to -1.When the target host completes the operation and allocates the memory, there are something wrong with the memory, causing the attacker gain the root privilege of the target host.

Thus, the atomic actions and the properties of an IMAP attack can be portrayed as follows.

\begin{enumerate}[(1)]
\item The value of ``literal" received by the target host is -1.
\item The target host completes the operation.
\item The target host allocates its memory space.
\item A memory error occurs in the target host.
\end{enumerate}

\paragraph{The formula for IMAP}~{}

According to the principles, we obtain the atomic propositions for the atomic actions and the properties in an Imap attack, as shown in Table \ref{tab23}. Therefore, the formula for an Imap attack can be given as follows

\begin{equation}
(literal.value =  - 1) \wedge \bigcirc (f;g;p)
\end{equation}

\subsubsection{Sendmail}

\paragraph{Principles}~{}

Sendmail is a general purpose email routing agent which supports many kinds of mail-transfer and delivery methods, mainly running on a variety of Unix/Linux operating systems. As a well-known free project, Sendmail is also an open source software. There exist some defects of codes for Sendmail in dealing with DNS requests, and it cannot effectively identify the returned parameters by a name server. Thus, by utilizing this vulnerability, a remote attacker can launch some buffer overflow attacks to a target host. As a result, he or she will finally get the privilege of Sendmail process. Sendmail cannot check perfectly the data returned by the name server when it attempts to use the mapping address with regard to the TXT query. Thus, the attacker can send a long string to the mail server, forging the returned information by the name server. Therefore, the buffer overflow attack occurs. If one carefully establishes the returned data, he or she can execute arbitrary commands on the system with the permission of the process of Sendmail. And in this way, the attacker can obtain the root privilege on the target host.

The main points are listed as follows.

If the target host receives a request for a Sendmail query, whose size is more than 256 bytes, one can determine that a Sendmail attack happens.

Thus, the atomic actions of a Sendmail attack can be portrayed as follows.
\begin{enumerate}[(1)]
\item The target host receives a request for a Sendmail query, and the number of its bytes is larger than 256.

\end{enumerate}

\paragraph{The formula for Sendmail}~{}

According to the principles mentioned above, we can give the atomic propositions for the atomic actions in a Sendmail attack, as shown in Table \ref{tab24}. Therefore, the formula for a Sendmail attack can be given as follows.

\begin{equation}
Attacked.receive.size > 256B
\end{equation}

\subsubsection{Xsnoop}

\paragraph{Principles}~{}

For a host server which has been infected with a Trojan, an attack can be easily launched from a client machine. An attacker lures the target host to run the program on the server, through some masquerading means. After the Trojans run, it will be hidden in the ’normal’ programs. These programs have been running silently since operating system is turned on. In other words, this type of Trojan program starts to run automatically with the start up of the system. The Trojan monitors all of the input from the keyboard of the target host, and it stores them on a local log. It is a key logger when the target host is offline. And it sends these records of log to the attacker when the target host is online.

The main points are listed as follows.

The Trojan program running on the target host stores the passwords entered from the keyboard in the local log. And it sends them to the attacker when the target host is online.

Thus, the atomic actions of an Xsnoop attack can be portrayed as follows.

\begin{enumerate}[(1)]
\item The Trojan running on the target host stores the passwords entered from the keyboard in the local log.
\item The Trojan sends the log including some passwords to the attacker.
\end{enumerate}

\paragraph{The formula for Xsnoop}~{}

According to the principles as stated above, we obtain the atomic propositions for the atomic actions in an Xsnoop attack, as shown in Table \ref{tab25}. Therefore, the formula for an Xsnoop attack can be given as follows.

\begin{equation}
attacked.password.save \wedge \bigcirc {\rm{(}}attacked.send.login)
\end{equation}

\section{ISIDMC}

\begin{figure}
		\centering
		\subfigure[for smurf attacks]{
			\begin{minipage}[b]{0.8\textwidth}
				\includegraphics[width=\textwidth]{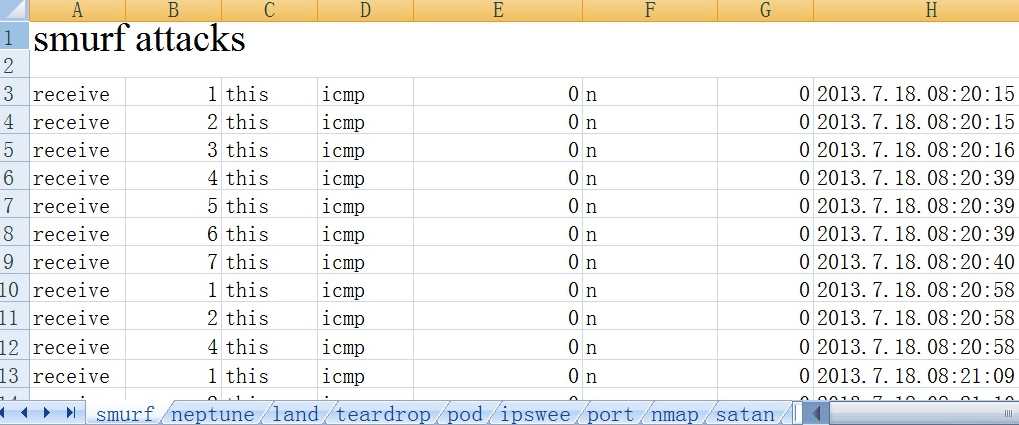} \\
				\label{fig1-1}
			\end{minipage}
		}
		
		\subfigure[ for warezdient attacks ]
		{
			\begin{minipage}[b]{0.8\textwidth}
				\includegraphics[width=\textwidth]{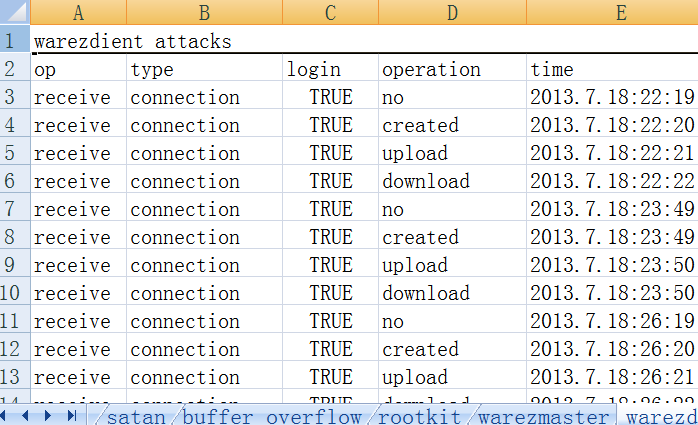} \\
				\label{fig1-2}
			\end{minipage}
		}

		\caption{some screenshots of the behavior-oriented intrusion set}
		\label{fig2}
	\end{figure}

The well-known KDDCUP provides a standard set for IDS benchmark experiments, which makes different IDS algorithms comparable on the unified platform. However, for a series of intrusion detection algorithms based on model checking, KDDCUP does not work, because the latter set provides a lot of connection-oriented attack records, and the former algorithm do work in accordance with the steps of attack. In other words, we need to construct our test sets due to the lack of available ones. To this end, we establish a more efficient record set, i.e., ISIDMC that contains thousands of attack records with changing patterns, (see Fig. \ref{fig2}). The process of constructing ISIDMC can be listed as follows.

At first, we approximately scale down the numbers of the twenty-four types of attacks in KDDCUP from thousands of thousands records to thousands ones.

And then, we map every connection-oriented record to its step-oriented version manually one by one. And we can do the same thing if several connection-oriented records mean one attack.

At last, we can append some changing patterns for every type of attacks.

It is obviously that ISIDMC can be used to compare the model-checking-based IDS algorithms since all of the records in the new set are step-oriented whereas these algorithms can only recognize step-oriented inputs. 

The key point of this paper is to compare the relative performance among the MC-based IDS algorithms. With the help of ISIDMC, we can do it for the ID algorithms based on temporal logic model checking in \cite{1}, \cite{2} and \cite{4} (see the next section). In this paper, we state that we conduct some benchmark tests because ISIDMC is originated from a Benchmark set called KDDCUP, and it is only existing one intrusion set that can be used to evaluate the model-checking-based IDS algorithms up to now. Compared with ISIDMC, KDDCUP cannot be employed to conduct the experiments for evaluating the MC-based IDS algorithms since none records in KDDCUP can make up an input of these algorithms.

Now, let us think about the following question: why is it possible that one can construct a benchmark set from the angle of behavior for MC-based IDS while attack patterns are changing continuously? On the one hand, some changed behaviors can be added to a benchmark set, although it cannot include all of changing continuously behaviors. On the other hand, the temporal logics include inherently numerous attack behaviors. Thus, an attack model expressed by the temporal logics can describe numerous attack behaviors. As a result, the model expressed by the temporal logics can deal with continuously changing behaviors, though the benchmark set cannot exhaust all the possibility.

For example, an attack may be performed as follows. At first, the attacker conducts an attack action a1. And then, he/she will conduct a normal action b1 in the following five seconds. At last, he/she will conduct an attack action a2 after ten seconds. However, this attack may be performed in the following way rather than the above way. At first, the attacker conducts an attack action a1. And then, he/she will conduct an attack action a2 after one second. It is obvious that the above two ways are equivalent if they are the same type of attacks. That is to say, the second way is a variant of the first one. 

For another example, an attack may be performed as follows. At first, the attacker conducts an attack action a1. And then, he/she will conduct a normal action b1. At last, he/she will conduct an attack action a2. However, this attack may be performed in the following way rather than the above way. At first, the attacker conducts an attack action a3 . And then, he/she will conduct an attack action a4. If both of the above two ways are consistent with the basic principle of the same type of attacks, they are equivalent, although their attack patterns are different. Therefore, the following formula can describe this type of attacks, no matter the difference of attack patterns, $a1 \wedge \bigcirc a2 \vee a3 \wedge \bigcirc a4$, where the sub formula $a1 \wedge \bigcirc a2$ describe the attack pattern 1 and the sub formula $a3\wedge \bigcirc a4$ describe the attack pattern 2, although the two attack patterns are belong to the same type of attacks.

What is more, the basic principle of this type of attacks indicates that the attacks’ principle must meet the property described below: an attack action a1 followed by another attack action a2, or an attack action a3 followed by another attack action a4. In other words, the attack pattern is changing, but its basic principle is constant. In fact, the number of these variants is numerous. Lucky for us, we can describe these numerous variants of the same type of attacks by using ONLY ONE temporal formula due to the constant basic principle of this type of attacks. In the example mentioned above, this formula is $a1\wedge \bigcirc a2 \vee a3 \wedge \bigcirc a4$. Thus, we model the twenty-four types of popular attacks by using ONLY the specific twenty-four temporal formulas, according to the twenty-four basic principles of twenty-four types of attacks, respectively.

Of course, as a misuse IDS technique, the model-checking-based algorithms cannot deal with any unknown attack. The user may need to improve the formula by adding a sub formulas, if a new pattern of the same type of attacks is found, so that the current formula can contain all of known patterns. 

As for the twenty-four types of attacks mentioned in this paper, the known attack patterns are contained by their basic principles. And we construct the model based on these basic principles RATHER THAN their specific attack patterns. Thus, we can say these models can deal with the known attack patterns. 

For example, if the basic principle of the smurf attacks was not as described in the above texts, it is no longer the smurf, although there are more and more new smurf attack modes changing and appearing continuously in the internet nowadays. This point guarantees that formula 1 gives the model of the smurf attacks, including its known various attack patterns.

In fact, the existing references has demonstrated the model-checking-based methods can deal with ``the changing attacks". Therefore, whether is it changing or not (i.e., the absolute abilities of these methods), this is not the research target of this paper since it is a solved problem (see \cite{1}\cite{2}\cite{3}\cite{4}\cite{6}\cite{7}). The goal in this paper is to make clear the relative abilities (instead of absolute abilities) of the existing model-checking-based methods. 

As for ``a new type of attacks", one can model it by using a new temporal formula. For example, the formula 21 has modeled the ``phf" attack, while the formula 22 has modeled the ``imap" attack. In other words, the different types of attacks are modeled by the different temporal formulas.  

This is about the models. As for the test set, the KDDCUP just provides some cases instead of full space of attack possibility for evaluating some IDS methods. Similarly, the set presented in this paper provides some typical cases on the changed attack behaviors instead of full space of changing continuously behaviors. Additionally, more cases can be added to the set in the further work. 

In a word, the following facts guarantee the new benchmark is reasonable and reliable.

\begin{enumerate}[(1)]
\item The existing benchmark set including KDDCUP cannot be employed to execute the MC-based IDS algorithms due to inconsistency of the I/O interface. Thus, our new benchmark is the ONLY ONE set which is compatible with the MC-based IDS algorithms in terms of I/O interface, to the best of our knowledge. As a result, ISIDMC is the ONLY ONE candidate when one compares the MC-based methods.
\item  ISIDMC originates from KDDCUP, as mentioned above. They are very good complementary for each other.
\item  Both the number of the types of attacks and the number of the attacks in ISIDMC further ensure that the new set is comprehensive, as shown in Table \ref{tab26}.
\item  The temporal formulas can express some changing attacks, as mentioned above.
\item  Fig. \ref{fig4} ensures that the ISIDMC-based comparisons of the MC-based methods are reasonable with regard to principle of algorithms.
\item  The experimental results in section V prove that these ISIDMC-based comparisons of the related algorithms are effective, operable and feasible in practices.
\end{enumerate}

\section{Simulation experiments}

An online LTL model checker called SPIN, an LTL-based IDS tool called Orchids and an ITL model checker can be seen in \cite{16}, \cite{17} an \cite{18}, respectively. The Fig. \ref{fig3} illustrates a screenshot of a running case of the RASL-based algorithm in \cite{4}.

We simulate these tools or algorithms via MATLAB. Our platform is: Intel i7-3770 CPU processor running at 3.40GHz, with 16 GB DDR3 memory on Windows 7. The existing network simulations do not suit our detection tasks. The reason is that there exists logical difference rather than physical difference among the three algorithms. The existing network simulations contain a lot of non-logical redundant information that will decline the detection efficiency, or even make us lose sight of the truth. Therefore, we do our simulation on MATLAB rather than the known network simulated tool.

\begin{figure}
		\centering
	
			\includegraphics[width= 0.5\textwidth]{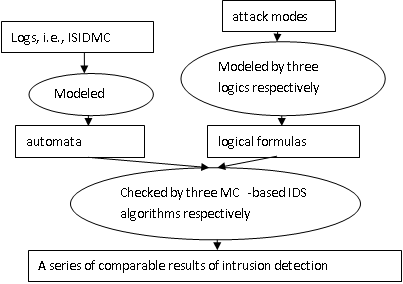}
			\caption{Principle of benchmark tests for the MC-based IDS algorithms}
			\label{fig4}

	\end{figure}

\begin{figure}
		\centering

			\includegraphics[width= 0.5\textwidth]{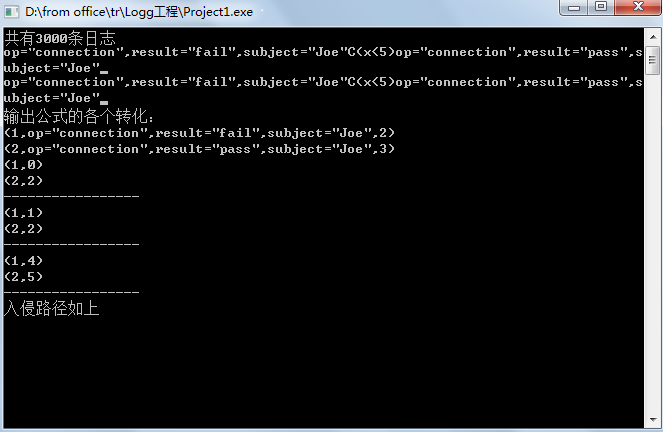}
			\caption{a screenshot of a running of the RASL-based algorithm}
			\label{fig3}
	
	\end{figure}

For every execution of each of these algorithms, there are two inputs. One is a formula in section 3, and the other is a record in ISIDMC. The principle of the detection is illustrated by the Fig. \ref{fig4}.

In fact, the framework presented in Fig. \ref{fig4}, i.e., the principles of the algorithms, were proposed by the existing references such as \cite{1}\cite{2}\cite{4} instead of this paper. All of the flow charts of the three algorithms are similar. The only difference is, each of them proposed a specific MC-based algorithms, whereas this paper replace it with ``three MC-based algorithms". 

In addition, Matlab is enough to the research goal of this paper, since the mission is explore the relative abilities rather than the absolute abilities for the existing model-checking-based methods. (BTW, some absolute abilities was explored in the existing references, such as \cite{1}, \cite{2}, \cite{3}, \cite{4}, \cite{6}, \cite{7}, \cite{8} and \cite{17}, and these works tell an user how about the absolute abilities of these algorithms with regard to chop-chop attacks, P-trace attacks and a part of DOS attacks). The difference of the relative abilities of the three existing algorithms only consist in the logical level. Furthermore, the matlab-based programs have simulated the main functions of the model checking algorithms with regard to the IDS topic. As a result, we can safety to say that Matlab is ENOUGH to the research goal of this paper.

In fact, all the simulated tools including Matlab can not exactly and totally present the results due to the abstract and its information loss. But the experimental results based on Matlab can verify one thing: the RELATIVE power of the different MC-based algorithms, due to the same effect on the abstract and information loss of the different algorithms.

In a word, in our simulated experiments, the SAME circumstance is used and the SAME set is input to the different algorithm in order to ensure the fairness of comparisons (another reason for using Matlab instead of the real IDS tools). No viewpoint is hold before the simulated experiments. According to the output set, we obtain the information on the different specialities of the different algorithms. This is the key idea of our simulated experiments.

\subsection{U2R attacks}
\begin{figure}
		\centering
	
			\includegraphics[width= 0.5\textwidth]{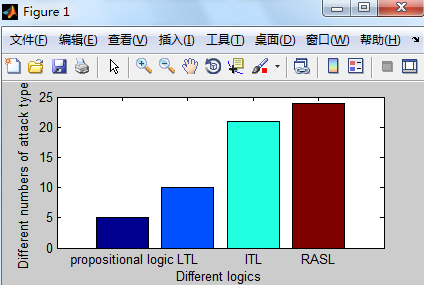}
			\caption{comparison: modeling abilities of the four logics}
			\label{fig5}
	
	\end{figure}

\begin{enumerate}[(1)]
\item {\bf The results}

The experimental results are shown in Fig. \ref{fig5}. It shows that, the stronger the expressive power of logic is, the stronger its modeling ability is.
\item {\bf The reasons}

The reasons for this situation include: compared with the propositional logic, LTL can describe temporal relationships among atomic actions in attack behaviors; compared with LTL, ITL can describe temporal relationships among non-atomic actions in attack behaviors and concurrent relationships among actions in attack behaviors; compared with ITL, RASL can describe the time constraint relationships among attack behaviors
\end{enumerate}

\subsection{a comparison of detection abilities between the methods}
\begin{figure}
		\centering

			\includegraphics[width= 0.5\textwidth]{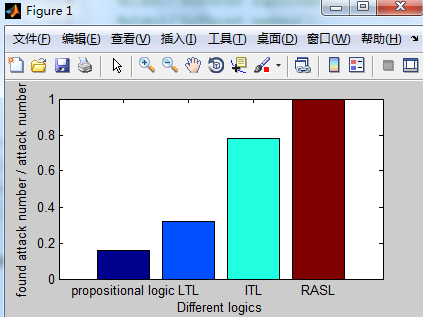}
			\caption{comparison: detection abilities of the four methods}
			\label{fig6}
		
	\end{figure}

As shown in Fig. \ref{fig6}, the stronger the expressive power of logic is, the more attacks the algorithms can detect. The reason is explained as above.

\subsection{pair-wise comparisons of the detection abilities between the algorithms}

\subsubsection{The propositional-logic-based algorithm VS the LTL-based algorithm}
\begin{figure}
		\centering
	
			\includegraphics[width= 0.5\textwidth]{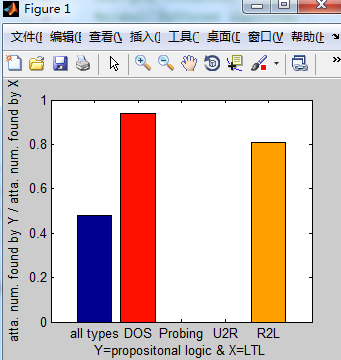}
			\caption{comparison: detection abilities (I) LTL Vs Propositional logic}
			\label{fig7}
	
	\end{figure}

\begin{enumerate}[(1)]
\item {\bf The results and the indication}

As shown in Fig. \ref{fig7}, the average detection rate of the latter algorithm reaches more than twice as high as the former one, for all of the twenty-four types of attacks. Especially, none of probing attacks and U2R attacks is found by the former algorithm.
\item {\bf The indication and the reason}

This result indicates that the LTL-based algorithm is more suitable for the detection of probing attacks and U2R attacks, compared with the propositional-logic-based algorithm due to the power of the logic employed by the LTL-based algorithm.

\end{enumerate}

\subsubsection{The LTL-based algorithm VS the ITL-based algorithm}

\begin{figure}
		\centering
	
			\includegraphics[width= 0.5\textwidth]{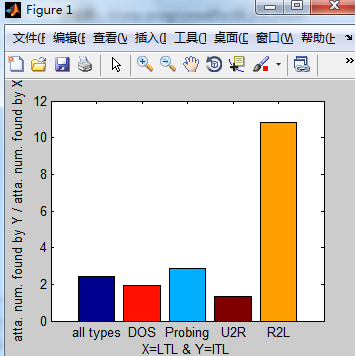}
			\caption{comparison: detection abilities (II) ITL Vs LTL}
			\label{fig8}

	\end{figure}

\begin{enumerate}[(1)]
\item {\bf The results}

The results are illustrated in Fig. \ref{fig8}. It shows that the latter algorithm finds more all types of attacks, especially for R2L attacks, than the former algorithm. This result indicates that the ITL-based algorithm is more suitable for the detection of R2L attacks, compared with the LTL-based algorithm.
\item {\bf The reason}

As shown in Section 3.4, there are a large number of interval relationships between actions in R2L attacks, which needs to be modeled by ITL formulas. This is why the ITL-based algorithm conduces a better performance.

\end{enumerate}

\subsubsection{The ITL-based algorithm VS the RASL-based algorithm}

\begin{figure}
		\centering
	
			\includegraphics[width= 0.5\textwidth]{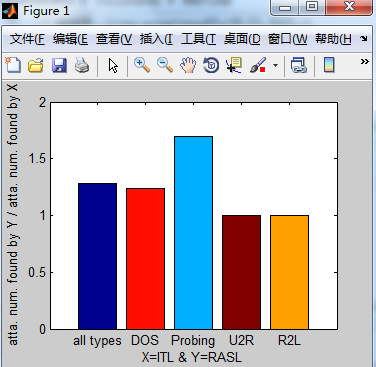}
			\caption{comparison: detection abilities (III) RASL Vs ITL}
			\label{fig9}

	\end{figure}

\begin{enumerate}[(1)]
\item {\bf The results and the indication}

The results are shown in Fig. \ref{fig9}. It illustrates that the latter algorithm can detect more attacks than the former algorithm, for DOS attacks and probing attacks. This result indicates that the RASL-based algorithm is more suitable for the detection of these attacks, compared with the ITL-based one.
\item {\bf The reason}

As shown in Section 3.1.6, Section 3.2.1 and Section 3.2.2, RASL formulas can express some time constraint relationships in mail bomb attacks, IP sweep attacks and port scan attacks. This is why the RASL-based algorithm shows a better performance.

\end{enumerate}

\subsection{Comparisons of the detection abilities for DOS, Probing, U2R and R2L}
\begin{figure}
\centering
			\includegraphics[width= 0.3\textwidth]{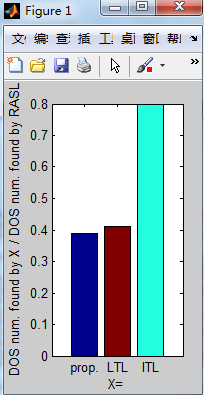}
			\caption{comparison: detection abilities (IV) for DOS attacks }
			\label{fig10}
	
	\end{figure}

\begin{figure}
		\centering
			\includegraphics[width= 0.3\textwidth]{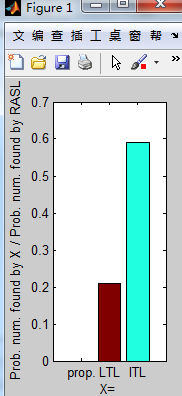}
			\caption{comparison: detection abilities (V) for Probing attacks}
			\label{fig11}
	\end{figure}

\begin{figure}
		\centering
	
			\includegraphics[width= 0.35\textwidth]{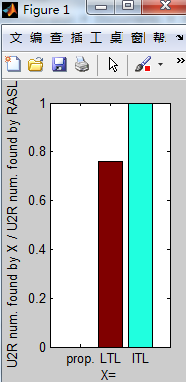}
			\caption{comparison: detection abilities (VI) for U2R attacks}
			\label{fig12}

	\end{figure}

\begin{figure}
\centering
\begin{minipage}[t]{0.4\textwidth}
\centering
\includegraphics{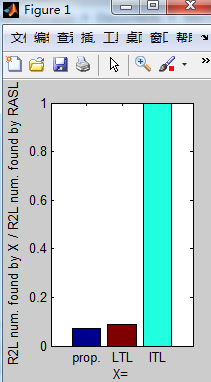}
\caption{comparison: detection abilities (VII) for R2L attacks}
\label{fig13}
\end{minipage}
\begin{minipage}[t]{0.4\textwidth}
\centering
\includegraphics{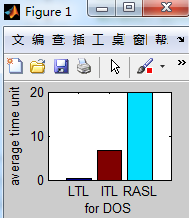}
\caption{ comparison: time consume (I) for DOS attacks}
\label{fig14}
\end{minipage}
\begin{minipage}[t]{0.8\textwidth}
\centering
\includegraphics{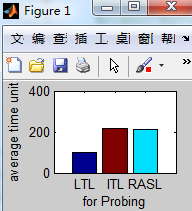}
\caption{  comparison: time consume (II) for Probing attacks  \\(The Propositional-logic-based method cannot find Probing attacks) }
\label{fig15}
\end{minipage}
\end{figure}

The results are illustrated in Fig. \ref{fig10}, Fig. \ref{fig11}, Fig. \ref{fig12}, and Fig. \ref{fig13}.

\begin{enumerate}[(1)]
\item {\bf For DOS attacks}

As shown in Fig. \ref{fig10}, the RASL-based algorithm can detect more DOS attacks than the ITL-based algorithm, whereas the ITL-based algorithm finds more DOS attacks than the LTL-based algorithm, due to the different expressive power of these logics.

\item {\bf For probing attacks}

See Fig. \ref{fig11}. The same thing happens due to the same reason, compared with the case of detecting DOS attacks. The only difference is observed as follows. The propositional-logic-based algorithm finds some DOS attacks and none of Probing attack since there exist some temporal relationships among actions in each type of Probing attacks.

\item {\bf For U2R attacks }

The same thing happens due to the same reason, compared with the case of detecting probing attacks. The only difference is observed as follows. As shown in Fig. \ref{fig12}, The ITL-based algorithm finds as many U2R attacks as the RASL-based algorithm since there exists none of time constraints among actions in each type of U2R attacks.

\item {\bf For R2L attacks}

See Fig. \ref{fig13}. The same thing happens due to the same reason, compared with the case of detecting U2R attacks. The only difference is that the propositional-logic-based algorithm finds some R2L attacks and none of U2R attack since there exists none of temporal relationships among actions in Sendmail attacks.
\end{enumerate}

\subsection{a comparison of time consumption of different algorithms}

\begin{figure}
\centering
\begin{minipage}[t]{0.4\textwidth}
\centering
\includegraphics{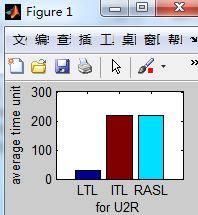}
\caption{comparison: time consume (III) for U2R attacks}
\label{fig16}
\end{minipage}
\begin{minipage}[t]{0.4\textwidth}
\centering
\includegraphics{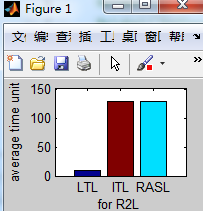}
\caption{  comparison: time consume (IV) for R2L attacks}
\label{fig17}
\end{minipage}
{\\ {} (The Propositional-logic-based method cannot find U2R attacks)}
\end{figure}

The results are shown in Fig. \ref{fig14}, Fig. \ref{fig15}, Fig. \ref{fig16} and Fig. \ref{fig17}. These figures illustrate the time cost in the scenarios of offline and online. For example, the offline version of the RASL-based algorithm takes 20 seconds to find a DOS attack hidden in a log. In comparison, the online version takes the equal time to find a DOS attack that occurs online. If these attacks occur more quickly, the online version of the RASL-based algorithm cannot deal with them.

Moreover, these figures show that the RASL-based algorithm takes much more than/the equal time with the ITL-based one for the detection, whereas the LTL-based algorithm takes much less time. This is due to the different time complexities of embedded model checking algorithms based on different temporal logics. The RASL model checking problem and the ITL model checking problems are high order exponential ones \cite{4}\cite{10}\cite{12}, whereas the LTL model checking problem is exponential one \cite{9}.

However, the specific complexities of intrusion detection algorithms are related to the specific formulas, when the different specific attacks are detected. The time cost varies due to the differences of formulas modeling the twenty-four types of attacks. In other words, the greater the number of embedment between the operator ``¬" and the operator ``;`` is, the higher the order of the exponent of the complexity is. This number reaches the maximum in formula 9 and formula 10. As a result, the ITL-based algorithm and the RASL-based one reach their maximum complexity, i.e., $O(2^{2n})$, when these two algorithms are employed to check some IP sweep attacks and/or Port Scan attacks. And the complexities of these two algorithms are lower than $O(2^{2n})$ when other twenty-two types of attacks are being detected. In short, the ITL-based algorithm and the RASL-based one have a worst complexity $O(2^{2n})$ when these  algorithms detect the popular DOS/Probing/U2R/R2L attacks. In the similar way, we find that the worst complexity of the LTL-based algorithm is $O(2^{n})$ when this algorithm detects these attacks.

\subsection{A comparison of the state space}
\begin{figure}
		\centering
			\includegraphics[width= 0.5\textwidth]{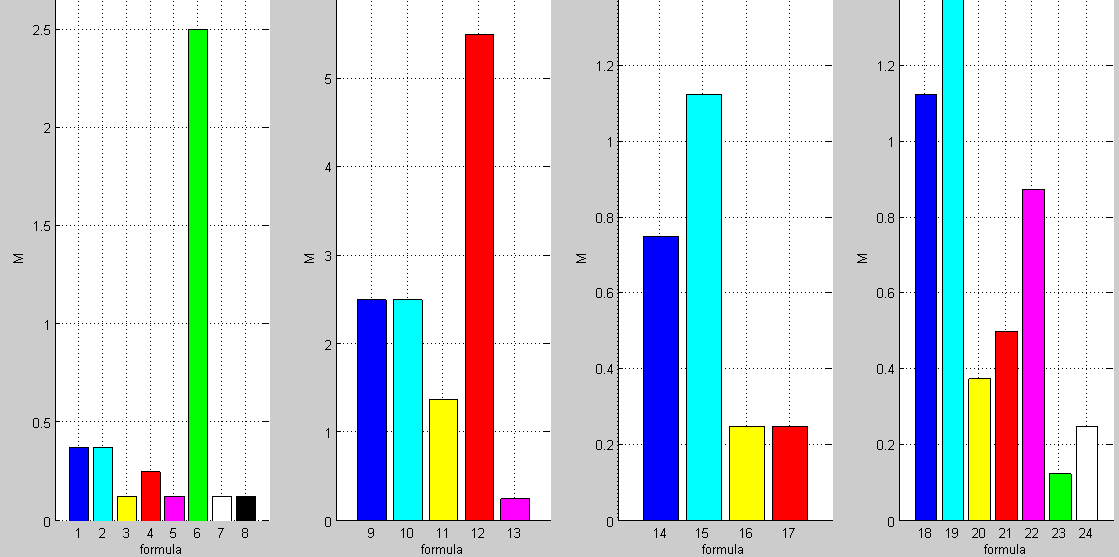}
			\caption{  comparison: memory consume that indicates the state space (from formula 1 to formula 24)  }
			\label{fig18}

	\end{figure}

The results are shown in Fig. \ref{fig18}. This figure illustrates the memory consumption of the MC-based algorithms with regard to the twenty-four types of attacks. These results indicate that the produced state space is acceptable when the existing MC-based methods are employed to detect the popular DOS/Probing/U2R/R2L attacks, since only simple temporal relationships occurs in these twenty-four models/formulas.

\section{Conclusion}
The experimental results in Section 5 illustrate the performance, in terms of detection abilities and time consumption, of the different IDS algorithms based on model checking for different types of attacks. As a result, we obtain the different recommended IDS algorithms based on model checking for the different types of attacks, as shown in Table \ref{tab26}. This is the conclusion of our benchmark experiments. It is beneficial to select proper MC-based algorithms in actual deployment of IDS.

\section*{Acknowledgment}
This work has been supported by National Natural Science Foundation of China under Grant No.U1204608

\begin{table}[!htb]
\centering
		\caption{INTUITIVE MEANING OF THE MAIN OPERATORS IN THE TEMPORAL LOGICS}
		\label{tab1}
		\begin{tabular}{|c|c|c|} \hline
		Operators &	Logic &	Illustrations \\ \hline
 		$\Box p$ &LTL & \raisebox{-0.9\height}{\includegraphics[width=0.3\textwidth]{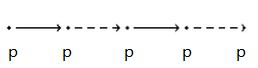}} \\ \hline

		$\diamondsuit p$ &LTL & \raisebox{-0.9\height}{\includegraphics[width=0.3\textwidth]{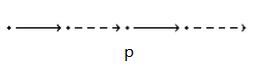}} \\ \hline

		${\varphi _1};{\varphi _2}$ &ITL & \raisebox{-0.9\height}{\includegraphics[width=0.3\textwidth]{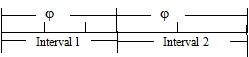}} \\ \hline

		${\varphi _1}*$ &ITL & \raisebox{-0.9\height}{\includegraphics[width=0.3\textwidth]{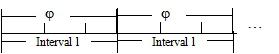}} \\ \hline
		
		${\varphi _1}{;_{x < 3}}{\varphi _2}$ & RASL &\raisebox{-0.9\height} {\includegraphics[width=0.3\textwidth]{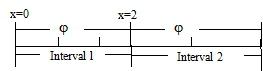}} \\ \hline

\end{tabular}

\end{table}

\begin{table}[!htb]
\centering
\caption{THE ATOMIC PROPOSITIONS IN THE FORMULA FOR SMURF ATTACKS AND THEIR MEANS}
		\label{tab2}
		\begin{tabular}{|c|c|c|} \hline
		Atomic propositions &	Meaning / definition \\ \hline

	$attacked.send$ & The attacked host sends a packet  \\ \hline
	$attacked.recieve$ & The attacked host receives a packet  \\ \hline

\end{tabular}
		
\end{table}

\begin{table} [!htb]
\centering
\caption{THE ATOMIC PROPOSITIONS IN THE FORMULA FOR NEPTUNE ATTACKS AND THEIR MEANS}
		\label{tab3}
		\begin{tabular}{|c|c|c|} \hline
		Atomic propositions &	Meaning / definition \\ \hline

	$attacked.recieve.SYN$ & The attacked host receives a SYN packet from a host  \\ \hline
	$attacked.send.SYNACK$ & The attacked host replies a SYNACK packet  \\ \hline
	$attacked.recieve.ACK$ & The attacked host receives an ACK packet \\ \hline
\end{tabular}
		
\end{table}

\begin{table}[!htb]
\centering
\caption{THE ATOMIC PROPOSITIONS IN THE FORMULA FOR LAND ATTACKS AND THEIR MEANS}
		\label{tab4}
		\begin{tabular}{|c|c|c|} \hline
		Atomic propositions &	Meaning / definition \\ \hline

	$attacked.recieve$ & The attacked host receives a packet  \\ \hline
	$p$ & Both source address and destination address are the attacked host address \\ \hline
\end{tabular}
		
\end{table}

\begin{table}[!htb]
\centering
\caption{THE ATOMIC PROPOSITIONS IN THE FORMULA FOR TEARDROP ATTACKS AND THEIR MEANS}
		\label{tab5}
		\begin{tabular}{|c|c|c|} \hline
		Atomic propositions &	Meaning / definition \\ \hline
	$m1.FragmentOffset$ & The fragment offset of the received packet 1 \\ \hline
	$m1.TotalLength$ & the total length of the received packet 1 \\ \hline
	$m2.FragmentOffset$ & The fragment offset of the received packet 2 \\ \hline
\end{tabular}
		
\end{table}

\begin{table}
\centering
\caption{THE ATOMIC PROPOSITION IN THE FORMULA FOR POD ATTACKS AND ITS MEANS}
		\label{tab6}
		\begin{tabular}{|c|c|c|} \hline
		Atomic propositions &	Meaning / definition \\ \hline

	$m.size > 65536$ & the size of the received packet is larger than 65536 byte  \\ \hline
\end{tabular}
		
\end{table}


	\begin{table}  
		\centering		
		\caption{THE ATOMIC PROPOSITIONS IN THE FORMULA FOR MAIL BOMB AND THEIR MEANS}
\label{tab15}
		\begin{tabular}{|c|c|} 
			\hline  
			Atomic propositions & Meaning / definition \\
			\hline  
			${p_i}$  & The target host receives the i-mail from the same IP address \\
			\hline
			$x$  & the time intervals among the ten records \\
			\hline
		\end{tabular}  
	\end{table}

	\begin{table}  
		\centering		
		\caption{THE ATOMIC PROPOSITIONS IN THE FORMULA FOR UDPSTORM AND THEIR MEANS}
		\label{tab16}
		
		\begin{tabular}{|c|c|} 
			\hline  
			Atomic propositions & Meaning / definition \\
			\hline  
			$attacked.receive.i$  & the target host receives the UDP packet from the host i in the subnet \\
			\hline
			$attacked.port$  & the port number of the target host \\
			\hline
			$i.udp.port$  & the port number of the attacker \\
			\hline
		\end{tabular}  
	\end{table}

	\begin{table}  
		\centering		
		\caption{THE ATOMIC PROPOSITIONS IN THE FORMULA FOR APACHE AND THEIR MEANS}
		\label{tab17}
		
		\begin{tabular}{|c|c|} 
			\hline  
			Atomic propositions & Meaning / definition \\
			\hline  
			$p$  & the target host receives an HTTP request \\
			\hline
			$attacked.receive.http.range$  & The HTTP request mentioned above contains a Range field \\
			\hline
			$attacked.receive.http.accept - encoding$  & The HTTP request mentioned above contains the Accept-Encoding field \\
			\hline
		\end{tabular}  
	\end{table}


\begin{table}
\centering
	\caption{THE ATOMIC PROPOSITIONS IN THE FORMULA FOR IP SWEEP AND THEIR MEANS}
		\label{tab7}
		\begin{tabular}{|c|c|c|} \hline
		Atomic propositions &	Meaning / definition \\ \hline

	$pi,i \in [1,10],i \in N$ & The event denoted by the i-th record  \\ \hline
\end{tabular}
	
\end{table}

\begin{table}
\centering
\caption{THE ATOMIC PROPOSITIONS IN THE FORMULA FOR PORT SCAN AND THEIR MEANS}
		\label{tab8}
		\begin{tabular}{|c|c|c|} \hline
		Atomic propositions &	Meaning / definition \\ \hline

	$qi,i \in [1,10],i \in N$ & The event denoted by the i-th record  \\ \hline
\end{tabular}
		
\end{table}

\begin{table}
\centering
	\caption{THE ATOMIC PROPOSITIONS IN THE FORMULA FOR NMAP AND THEIR MEANS}
		\label{tab9}
		\begin{tabular}{|c|l|} \hline
		Atomic propositions or sub-formulas &	  \makecell[c]{Meaning / definition} \\ \hline
	 $attacked.recieve.ICMP - echo - request$ & \multirow{4}{9cm}{Host discovery: sending four kinds of detection packets---ICMP echo request, a TCP SYN packet to port 443, a TCP ACK packet to port 80, an ICMP timestamp request}   \\  \cline{1-1}
     $attacked.recieve.TCPSYN$   & \\  \cline{1-1}
$attacked.recieve.TCPACK$  &  \\  \cline{1-1}
$attacked.recieve.ICMP - timestamp - request$  & \\ \cline{1-1} \hline

  \makecell[l]{\\$attacked.recieve.SYN$ \\ {}}  &  \multirow{5}{9cm}{Port Scan: \\
There are six kinds of scan mode ---TCP SYN scanning: \\
$attacked.recieve.SYN$ \\
TCP ACK scanning: \\
$attacked.recieve.ACK$\\
TCP FIN scanning:\\
$attacked.recieve.TCPFIN$\\
TCP Xmas scanning:\\
 $attacked.recieveTCP.flags.FINURGPUSH=1$ \\
TCP NULL scanning:\\
$attacked.recieve.TCP.flags = 0$\\
UDP scanning:\\
$attacked.port.UDP.recieve.ICMP$ 
}   \\  \cline{1-1}
\makecell[l]{\\ $attacked.recieve.ACK$\\ {} } & \\  \cline{1-1}

\makecell[l]{\\ $attacked.recieve.TCPFIN$\\ {} } & \\  \cline{1-1}

\makecell[l]{\\  $attacked.recieve.TCP.flags.FINURGPUSH$\\ {} } & \\  \cline{1-1}

\makecell[l]{\\  $attacked.recieve.TCP.flags$\\ {} } & \\  \cline{1-1}

\makecell[l]{\\  $attacked.port.UDP.recieve.ICMP$ \\ {} } & \\  \hline

\makecell[l]{  \\ $Exclusionlist.check.status.open$ \\{}} & \multirow{4}{9cm}{Version detection:\\Checking whether the port of status open is in the exclusion of the port list:\\
$Exclusionlist.check.status.open$ \\
Checking whether the port of status open filter is in the exclusion of the port list:
$Exclusionlist.check.status.openfiltered$ If it is the TCP port, try to establish a TCP connection:\\
$port.TCP.TCPconnect$
If it is the UDP port, then use the nmap-services-probes in the probe packets to match:
$port.UDP.recieve.nmapserviecesprobes$}  \\  \cline{1-1}

\makecell[l]{\\  $Exclusionlist.check.status.openfiltered$ \\{} } &  \\  \cline{1-1}
 
\makecell[l]{ \\ $port.TCP.TCPconnect$ \\{}} &  \\  \cline{1-1}

\makecell[l]{  \\ $port.UDP.recieve.nmapserviecesprobes$\\{} } &  \\  \hline

\makecell[l]{$port.open.recieve.TCP$} &  \makecell[l]{Operating system detection:\\Sending TCP packet to open port:\\ $port.open.recieve.TCP$ \\ Sending UDP packet to open port: \\
$port.open.recieve.UDP$\\Sending ICMP packet to open port:\\ $port.open.recieve.ICMP$ \\ Sending TCP packet to closed port:\\ $port.closed.recieve.TCP$ \\ Sending UDP packet to closed port: \\$port.closed.recieve.UDP$\\ Sending ICMP packet to closed port:\\ $port.closed.recieve.ICMP$} \\ \hline

\end{tabular}
	
\end{table}

	\begin{table}  
		\centering		
		\caption{ THE SUB-FORMULAS IN THE FORMULA FOR SATAN AND THEIR MEANS}
		\label{tab10}
		
		\begin{tabular}{|c|c|} 
			\hline  
			Atomic propositions or sub-formulas & Their means \\
			\hline  
			$attacked.nslookuped\_program$  & An ITL formula for Ns lookup scan  \\
			\hline
			$attacked.portmapped\_program$  & An ITL formula for Portmap scan  \\
			\hline
			$attacked.showmount\_program$  & An ITL formula for Show mount scan \\
			\hline
			$attacked.fingered.scanned\_program$  & An ITL formula for Finger scan \\
			\hline
			$attacked.TCP.scanned\_program$  & An ITL formula for TCP scan  \\
			\hline
			$attacked.UDP.scanned\_program$  & An ITL formula for UDP scan \\
			\hline
			$attacked.activeservices.scanned\_program$  & An ITL formula for some active service scan \\
			\hline
		\end{tabular}  
	\end{table}
	

	\begin{table}  
		\centering		
		\caption{THE PROPOSITIONS IN THE FORMULA FOR MSCAN AND THEIR MEANS}
		\label{tab18}
		
		\begin{tabular}{|c|c|} 
			\hline  
			propositions & Meaning / definition \\
			\hline  
			$attacked.receieve.SYN \wedge port = 113$  & TCP SYN scanning \\
			\hline
			$attacked.receieve.SYN \wedge port = 21$  & TCP connect scanning \\
			\hline
			$attacked.receieve.SYN \wedge port = 389$  & TCP ACK scanning \\
			\hline
			$attacked.receieve.SYN \wedge port = 443$  & TCP FIN scanning \\
			\hline
			$attacked.receieve.TCP.flags.FINURGPUSH = 1 \wedge port = 443$  & TCP Xmas scanning \\
			\hline
			$attacked.receieve.TCP.flags = 0 \wedge port = 443$  & TCP NULL scanning \\
			\hline
			$attacked.port.UDP.receive.ICMP \wedge port = 971$  & UDP scanning \\
			\hline
			$attacked.receieve.TCP.{\rm{ }}portmapper \wedge port = 135$  & \makecell{The port mapper provides a \\ running RPC service} \\
			\hline
			$attacked.receieve.nfsd.exportfs$  & NFSD provides a table of contents \\
			\hline
			$attacked.receieve.samba \vee attacked.receieve.netbios$  & \makecell{Samba or netbios provides \\ a shared table of contents} \\
			\hline
			$attacked.receieve.finger \wedge port = 79$  & \makecell{use finger to determine whether or not \\ the default account exists} \\
			\hline
			$\begin{array}{l}
			(port.open.receive.TCP \vee port.open.receive.UCP\\
			\vee port.open.receive.ICMP) \wedge (port.closed.receive.TCP\\
			\vee port.{\rm{ }}closed.receive.UCP \vee port.closed.ICMP)
			\end{array}$  & detection of operating systems \\
			\hline
		
		\end{tabular}  
	\end{table}

	\begin{table}  
		\centering		
		\caption{THE ATOMIC PROP. OR SUB-FORMUL. IN THE FORMULA FOR BUFFER OVERFLOW AND THEIR MEANS}
		\label{tab11}
		
		\begin{tabular}{|c|c|} 
			\hline  
			Atomic propositions or sub-formulas & Meaning / definition \\
			\hline  
			$attacked.recieve.string$  & The attacked host receives the string.  \\
			\hline
			$code.\bmod ified$  & The parameters of the existing codes have been modified.  \\
			\hline
			$string\_program$  & \makecell{An ITL formula for the received string containing attack sequence \\ which can run on the attacked hardware platforms.} \\
			\hline
			$code.execute\_program$  & An ITL formula for the modified codes. \\
			\hline
		\end{tabular}  
	\end{table}
	
	\begin{table}  
		\centering		
		\caption{THE SUB--FORMUL. IN THE FORMULA FOR ROOTKIT AND THEIR MEANS}
		\label{tab12}
		
		\begin{tabular}{|c|c|} 
			\hline  
			Atomic propositions or sub-formulas & Meaning / definition \\
			\hline  
			$code.\bmod ified\_program$  & An ITL formula for replacing the system monitoring program.  \\
			\hline
			$sys\log .\bmod ified\_program$  & An ITL formula for modifying system log.  \\
			\hline
			$sys\log .delete\_program$  & An ITL formula for deleting system log. \\
			\hline
		\end{tabular}  
	\end{table}

	\begin{table}  
		\centering		
		\caption{THE ATOMIC PROPOSITIONS IN THE FORMULA FOR HTTPTUNNEL AND THEIR MEANS}
		\label{tab19}
		
		\begin{tabular}{|c|c|} 
			\hline  
			Atomic propositions & Meaning / definition \\
			\hline  
			$client.htc$  & the htc runs on client host \\
			\hline
			$packets.port.i$  & The port number of the packet \\
			\hline
			$client.send.http$  & The htc sends the encapsulated HTTP request through the firewall to the outside network \\
			\hline
		\end{tabular}  
	\end{table}


	\begin{table}  
		\centering		
		\caption{THE ATOMIC PROPOSITIONS IN THE FORMULA FOR XTERM AND THEIR MEANS}
		\label{tab20}
		
		\begin{tabular}{|c|c|} 
			\hline  
			Atomic propositions & Meaning / definition \\
			\hline  
			$p$  & The target host receives the given malicious codes \\
			\hline
			$Attack.receive.escape.threshold$  & The target host received the window threshold of the Escape command \\
			\hline
			$Banner.modifed.FTP$  & Banner information is modified on FTP \\
			\hline
			$Banner.modifed.TELNET$  & The banner information is modified on TELNET \\
			\hline
			$Syslog.modifed$  & The target host's system log has been modified \\
			\hline
			$Syslog.Symlinked$  & there exists a symbolic link between the two permissions \\
			\hline
		\end{tabular}  
	\end{table}

	\begin{table}  
		\centering		
		\caption{THE SUB--FORMUL. IN THE FORMULA FOR WAREZMASTER AND THEIR MEANS}
		\label{tab13}
		
		\begin{tabular}{|c|c|} 
			\hline  
			sub-formulas & Meaning / definition \\
			\hline  
			$account.guest.\log in\_program$  & \makecell{An ITL formula for the process of the attacker using the guest \\account to login to the server}  \\
			\hline
			$hiddendirectory.created\_program$  & An ITL formula for the process of the attacker creating the hidden dictionary.  \\
			\hline
			$uploadwarez\_program$  &An ITL formula for the process of the attacker uploading ware-z to server  \\
			\hline
		\end{tabular}  
	\end{table}
	
	\begin{table}  
		\centering		
		\caption{THE SUB--FORMUL. IN THE FORMULA FOR WAREZCLIENT AND THEIR MEANS}
		\label{tab14}
		
		\begin{tabular}{|c|c|} 
			\hline  
			sub-formulas & Meaning / definition \\
			\hline  
			$downloadwarez\_program$  & An ITL formula for the process of the user downloading ware-z  \\
			\hline
		\end{tabular}  
	\end{table}

	
	\begin{table}  
		\centering		
		\caption{THE ATOMIC PROPOSITIONS IN THE FORMULA FOR FTP\_WRITE AND THEIR MEANS}
		\label{tab21}
		
		\begin{tabular}{|c|c|} 
			\hline  
			Atomic propositions & Meaning / definition \\
			\hline  
			$attacked.{\rm{ }}create.file$  & create a new file on the home directory on the FTP server \\
			\hline
			$file.p$  & the suffix of the new file \\
			\hline
			$attacked.open.rlogin$  & The FTP server opens the process of rlogin \\
			\hline
		\end{tabular}  
	\end{table}
	
	\begin{table}  
		\centering		
		\caption{THE ATOMIC PROPOSITIONS IN THE FORMULA FOR PHF AND THEIR MEANS}
		\label{tab22}
		
		\begin{tabular}{|c|c|} 
			\hline  
			Atomic propositions & Meaning / definition \\
			\hline  
			$attacked.receive.http$  & The target host receives the same HTTP request \\
			\hline
			$p$  & the HTTP request contains``$\verb|\|  n$" and hexadecimal 0x0a \\
			\hline
			$attacked.xterm$  & The target server executes an Xterm command \\
			\hline
			$attacked.telnet$  & The target server generates a reverse telnet connection \\
			\hline
			$telnet.port$  & The port number of the reverse telnet connection \\
			\hline
		\end{tabular}  
	\end{table}

	\begin{table}  
		\centering		
		\caption{THE ATOMIC PROPOSITIONS IN THE FORMULA FOR IMAP AND THEIR MEANS}
		\label{tab23}
		
		\begin{tabular}{|c|c|} 
			\hline  
			Atomic propositions & Meaning / definition \\
			\hline  
			$literal.value$  & The value of the received ``literal" \\
			\hline
			$f$  & The target host complete the operation \\
			\hline
			$g$  & The target host allocates memory space \\
			\hline
			$p$  & The target host has a memory error \\
			\hline
		\end{tabular}  
	\end{table}
	
	\begin{table}  
		\centering		
		\caption{THE ATOMIC PROPOSITIONS IN THE FORMULA FOR SENDMAIL AND THEIR MEANS}
		\label{tab24}
		
		\begin{tabular}{|c|c|} 
			\hline  
			Atomic propositions & Meaning / definition \\
			\hline  
			$Attacked.receive.size$  & The target host receives a request for a Sendmail query \\
			\hline
		\end{tabular}  
	\end{table}
	
	\begin{table}  
		\centering		
		\caption{THE ATOMIC PROPOSITIONS IN THE FORMULA FOR XSNOOP AND THEIR MEANS}
		\label{tab25}
		
		\begin{tabular}{|c|c|} 
			\hline  
			Atomic propositions & Meaning / definition \\
			\hline  
			$attacted.password.save$  & The Trojan running on the target host save the password in the local log \\
			\hline
			$attacted.send.login$  & The Trojan sends the log including some passwords to the attacker \\
			\hline
		\end{tabular}  
	\end{table}

\begin{table}
\centering
\caption{ VARIOUS TYPES OF ATTACKS AND THEIR SUITABLE ALGORITHMS}
\label{tab26}
\begin{tabular}{|c|c|c|c|c|c|}
\hline
\makecell[c]{Types of \\ attacks} & \makecell[c]{amount for \\ each type of \\ attacks in the\\ Benchmark set} & \makecell[c]{Logic and \\ Algorithms} & \makecell[c]{Types of \\ attacks} & \makecell[c]{amount for \\ each type of \\attacks in the \\ Benchmark set} & \makecell[c]{Logic and \\ Algorithms} \\ \hline
Smurf            & 100                                                  & ITL                  & Mscan            & 100                                                  & LTL                  \\ \hline
Neptune          & 100                                                  & LTL                  & Buffer Overflow  & 22                                                   & ITL                  \\ \hline
Land             & 9                                                    & Propositional logic  & Rootkit          & 13                                                   & ITL                  \\ \hline
Teardrop         & 12                                                   & LTL                  & Httptunnel       & 100                                                  & LTL                  \\ \hline
Pod              & 87                                                   & Propositional logic  & Xterm            & 13                                                   & ITL                  \\ \hline
Mail bomb        & 100                                                  & RASL                 & Warezmaster      & 100                                                  & ITL                  \\ \hline
Udpstorm         & 2                                                    & Propositional logic  & Warezclient      & 100                                                  & ITL                  \\ \hline
Apache           & 100                                                  & Propositional logic  & Ftp write        & 3                                                    & ITL                  \\ \hline
IP sweep         & 100                                                  & RASL                 & Phf              & 2                                                    & ITL                  \\ \hline
Port scan        & 100                                                  & RASL                 & Imap             & 1                                                    & ITL                  \\ \hline
Nmap             & 84                                                   & ITL                  & Sendmail         & 17                                                   & Propositional logic  \\ \hline
Satan            & 100                                                  & ITL                  & Xsnoop           & 4                                                    & LTL                  \\ \hline
\end{tabular}
\end{table}
	
\end{document}